\documentclass[aps,prb,showpacs,twocolumn,floats,epsfig,pdflatex]{revtex4}
\usepackage{epsfig}
\usepackage{amsmath}
\usepackage{amsfonts}
\usepackage{graphicx}
\usepackage{amssymb}
\usepackage{amsbsy}
\usepackage{array}
\usepackage{color}

\newcolumntype{?}[1]{!{\vrule width #1}}
\makeatletter
\def\hlinewd#1{%
  \noalign{\ifnum0=`}\fi\hrule \@height #1 \futurelet
   \reserved@a\@xhline}
\makeatother

\begin{document}

\title {Magnetization induced dynamics of a Josephson junction coupled to a nanomagnet}

\author{Roopayan Ghosh$^{(1)}$, Moitri Maiti$^{(2)}$, Yury M. Shukrinov,$^{(2,3)}$ and K. Sengupta$^{(1)}$}

\affiliation{$^{(1)}$Theoretical Physics Department, Indian
Association for the Cultivation of Science, Jadavpur,
Kolkata-700032, India.\\
$^{(2)}$ BLTP, JINR, Dubna, Moscow region, 141980, Russia \\
$^{(3)}$ Dubna State University, Dubna, Russian Federation.}

\date{\today}

\begin{abstract}

We study the superconducting current of a Josephson junction (JJ)
coupled to an external nanomagnet driven by a time dependent
magnetic field both without and in the presence of an external AC
drive. We provide an analytic, albeit perturbative, solution for the
Landau-Lifshitz (LL) equations governing the coupled JJ-nanomagnet
system in the presence of a magnetic field with arbitrary
time-dependence oriented along the easy axis of the nanomagnet's
magnetization and in the limit of weak dimensionless coupling
$\epsilon_0$ between the JJ and the nanomagnet. We show the
existence of Shapiro-like steps in the I-V characteristics of the JJ
subjected to a voltage bias for a constant or periodically varying
magnetic field and explore the effect of rotation of the magnetic
field and the presence of an external AC drive on these steps. We
support our analytic results with exact numerical solution of the LL
equations. We also extend our results to dissipative nanomagnets by
providing a perturbative solution to the Landau-Lifshitz-Gilbert
(LLG) equations for weak dissipation. We study the fate of
magnetization-induced Shapiro steps in the presence of dissipation
both from our analytical results and via numerical solution of the
coupled LLG equations. We discuss experiments which can test our
theory.

\end{abstract}


\maketitle

\section{Introduction}
\label{intro}

The physics of Josephson junctions (JJs) has been the subject of
intense theoretical and experimental endeavor for decades
\cite{likharev1}. The interest in the physics of such JJs has
received renewed attention in recent years in the context of
Majorana modes in unconventional superconductors
\cite{kit1,kwon1,jay1,gil1}. Indeed, it has been theoretically
predicted \cite{kit1,kwon1}and experimentally observed
\cite{rokhin1} that such junctions may serve as a test bed for
detection of Majorana end modes in unconventional superconductors.
It has been shown that the presence of such end modes lead to
fractional Josephson effect \cite{kit1,kwon1} and results in the
absence of odd Shapiro steps \cite{shapiro1} when such junctions are
subjected to an external AC drive \cite{evenlit,kiril1,commentsub}.

Recently molecular nanomagnets have been studied as potential
candidates for qubit realization owing to their long magnetization
relaxation times at low temperatures \cite{time1}. Such a
realization is expected to play a central role in several aspects of
quantum information processing \cite{loss1} and spintronics using
molecular magnets \cite{ref1,ref2}. These systems have potential for
high-density information storage and are also excellent examples of
finite-size spin systems which are promising test-beds for
addressing several phenomena in quantum dynamics viz.
quantum-tunneling of the magnetization \cite{tunnel}, quantum
information \cite{qi}, entanglement \cite{ent1} to name a few. The
study of the spin dynamics of the nanomagnets is a crucial aspect of
all such studies. One way to probe such dynamics is to investigate
the spin response in bulk magnets using inelastic neutron scattering
and subsequent finite-size extrapolation to obtain the inelastic
neutron scattering spectra for a single molecule \cite{spin1}.
Other, more direct, methods include determination of the real-space
dynamical two-spin correlations in high-quality crystals of
nanomagnets \cite{spin2} and transport measurements through
nanomagnets \cite{spin3}. Another probe of such magnetization
dynamics, which we shall focus on in the present study, involves
interaction of the nanomagnet with a JJ; the modulation of the
Josephson current and the Shapiro steps in the current-voltage (I-V)
characteristics is known to bear signature of the nature of the spin
precession of the nanomagnet \cite{spin4}.

The physics of such a JJ-nanomagnet system has therefore received
significant attention both theoretically and experimentally. It has
been theoretically studied in Refs.\ \onlinecite{kulik1,bula1},
where the effects of spin-flip of the nanomagnet on the Josephson
current was charted out. More recently, several theoretical studies
have been carried out on a variety of aspects of such systems
including effect of superconducting correlations on the spin
dynamics of the nanomagnet \cite{nuss1}, the influence of spin-orbit
coupling of a single spin on the Josephson current \cite{nazarov1},
and the effect of deposition of single magnetic molecules on
superconducting quantum interference devices(SQUIDs) made out of
such junctions \cite{delan1}. Another interesting phenomenon which
has been widely studied in this context is magnetization switching
\cite{cai1,buzhdin1,yury1} which constitutes magnetization reversal
of the nanomagnet by a externally driven JJ. In addition, the
possibility of a Josephson current to induce exchange interaction
between ferromagnetic layers has been studied in Ref.\
\onlinecite{xav1}. Furthermore, the dynamics of both JJs with
misaligned ferromagnetic layers and those coupled to single or
multiple ferromagnetic layers have also been studied numerically
\cite{linder1}. It has also been shown that the presence of a single
classical spin of a molecular magnet precessing under the action of
a constant magnetic field coupled to a JJ may lead to generation of
finite spin current whose polarization axis rotates with same
frequency as the classical spin \cite{holm1}. Such theoretical
studies were complemented by experimental work on these systems
\cite{wernsdorfer1}. More recently, magnetization reversal of a
single spin using a JJ subjected to a static field and a weak
linearly polarized microwave radiation has been demonstrated in
Ref.\ \onlinecite{thirion1}. The possibility of the presence of
Shapiro-like steps in the I-V characteristics of such coupled
JJ-nanomagnet for constant applied magnetic field has also been
pointed out in Ref.\ \onlinecite{cai1}. However, to the best of our
knowledge, most of these studies do not provide any analytic
treatment of the coupled JJ-nanomagnet system even at a classical
level where they are known to be governed by the
Landau-Lifshitz-Gilbert (LLG) equations \cite{gil1}. Moreover, the
current-voltage (I-V) characteristics of a JJ in the presence of a
nanomagnet with time-dependent magnetic fields and in the presence
of external AC drive has not been studied systematically so far.

In this work we study a JJ coupled to a nanomagnet with a fixed
easy-axis anisotropy direction (chosen to be $\hat y$) in the
presence of an arbitrary time dependent external magnetic field
along $\hat y$. For nanomagnets with weak anisotropy, we find an
analytic perturbative solution to the coupled Landau-Lifshitz (LL)
equations in the limit of weak coupling between the nanomagnet and
the JJ. Using this solution, we show that a finite DC component of
the supercurrent, leading to Shapiro-like steps in the I-V
characteristics of a voltage-biased JJ, can occur, in the absence of
any external radiation, for either a constant or a periodically time
varying magnetic field. Our theoretical analysis provides exact
analytic results for the position of such steps. We study the
stability of these steps against change in the direction of the
applied magnetic field and increase of the dimensionless coupling
strength $\epsilon_0$ between the JJ and the nanomagnet. We also
provide a detailed analysis of the fate of this phenomenon in the
presence of an external AC drive and demonstrate that the presence
of such a drive leads to several new fractions (ratio between the
applied DC voltage and the drive frequency) at which the
supercurrent develops a finite DC component leading to Shapiro steps
in the I-V characteristics. We support our analytical results with
numerical study of the systems which allows exact, albeit numerical,
investigation of the dynamics of the coupled JJ and nanomagnet
system. We also extend our study to systems with dissipation via
perturbative analytic and exact numerical solution of the coupled
LLG equations and study the behavior of the steps with increasing
dissipation. Finally, we discuss experiments which may test our
theory, discuss the significance of our results in context of
junctions of unconventional superconductors hosting Majorana end
modes, and point out the distinction between voltage and current
biased junctions in the present context. We note that the analytic
solution of the LLG equations that we present here as well as the
presence of additional, more robust, Shapiro steps for periodically
time-varying magnetic fields have not been reported so far; thus our
work may serve as an ingredient for development of new detectors for
magnetization of a nanomagnet in coupled JJ-nanomagnet systems.

\begin{figure}[ht]
\centering
\includegraphics[scale=0.35]{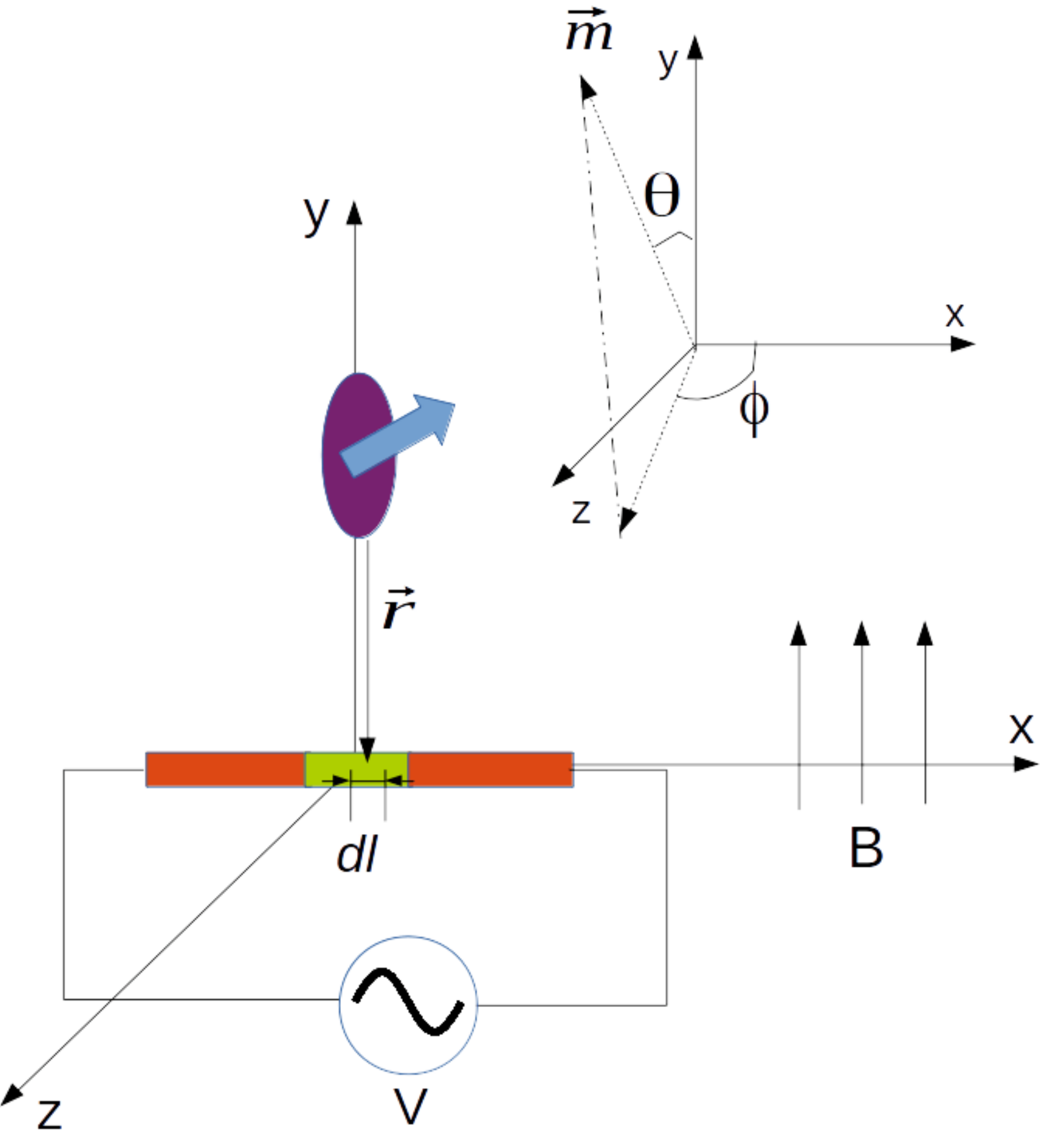}
\caption{A schematic representation of the JJ-nanomagnet system
showing the position of the nanomagnet (shown schematically by the
filled oval and an arrow representing the direction of its
instantaneous magnetization) and the JJ (orthogonal to the direction
of the magnetic field). The inset shows the angles $\theta$ and
$\phi$ used to specify the direction of the nanomagnet's
magnetization $\vec m = \vec M/|\vec M|$.} \label{fig1}
\end{figure}

The plan of the rest of this work is as follows. In Sec.\
\ref{ansol}, we provide the analytic solution of the LL equations
governing the coupled nanomagnet and JJ. This is followed by an
analogous solution for LLG equations describing the coupled
JJ-nanomagnet system in the presence of dissipation in Sec.\
\ref{gilres}. Next, in Sec.\ \ref{numsol}, we back the analytical
results with exact numerics and discuss details of Shapiro-step like
features in the I-V of the JJ for constant or periodic magnetic
field. Finally, we chart out our main results, discuss experiments
which can test our theory, and conclude in Sec.\ \ref{dissec}.

\section{Formalism and analytical solution}
\label{ansol}

In this section, we obtain analytic solution to the LL equations for
the weakly coupled JJ-nanomagnet system. We shall sketch the general
derivation of our result for arbitrary time-dependent magnetic field
in Sec.\ \ref{genres} and then apply these results to demonstrate
the existence of Shapiro-like steps for constant and periodic
magnetic fields in Sec.\ \ref{pecresults}. The extension of these
results for dissipative magnets will be charted out in Sec.\
\ref{gilres}.

\subsection{Perturbative solution}
\label{genres}

The coupled JJ-nanomagnet system is schematically shown in Fig.\
\ref{fig1}. In what follows we consider a JJ along $\hat x$ and
choose the easy axis of the nanomagnet along $\hat y$; the radius
vector $\vec r$ between the nanomagnet and the JJ thus lies in the
$x-y$ plane as shown in Fig.\ \ref{fig1}. The energy functional
governing the JJ and the nanomagnet is given by \cite{cai1}
\begin{eqnarray}
E &=& E_1 + E_2 \nonumber\\
E_1 &=& -K M_y^2 - M_y B(t), \quad E_2 = -E_J \cos \gamma,
\label{en1}
\end{eqnarray}
where $K>0$ denotes the magnetization anisotropy constant, $\vec
B(t)\parallel {\hat y}$ is the external magnetic field which can
have arbitrary time dependence, and $E_J$ is the Josephson energy of
the junction. The phase difference $\gamma$ across the junction is
given by

\begin{eqnarray}
\gamma(t) &=&  \gamma_0(t) + \gamma_1(t), \nonumber\\
\gamma_0(t) &=& \gamma_{00}+ \frac{2e}{\hbar}
\int^t V_0(t') dt' = \gamma_{00} + \omega_0' \int^t g(t') dt' \nonumber\\
\gamma_1(t) &=& - \frac{2 \pi}{\Phi_0} \int {\vec dl} \cdot {\vec
A(\vec r)}, \label{phasedef}
\end{eqnarray}
where $\gamma_{00}$ is the intrinsic DC phase of the JJ, $\gamma_0$
is the phase generated by the external voltage, $V_0(t)=V_0 g(t)$ is
the applied voltage across it, $\omega_0'=2eV_0/\hbar$ is the
Josephson frequency of the junction, $g(t)$ is a dimensionless
function specifying the time dependence of the applied voltage,
$\Phi_0=  h c/2e$ is the flux quantum, $h= 2 \pi \hbar$ with $\hbar$
being the Planck constant, $e$ is the charge of an electron, and $c$
is the speed of light. The vector potential ${\vec A(\vec r)}$ is
given by
\begin{eqnarray}
{\vec A(\vec r,t)} &=& \mu_0 (\vec M(t)  \times \vec r)/(4 \pi |\vec
r|^3). \label{vecpot}
\end{eqnarray}
Note that in our chosen geometry, as shown in Fig.\ \ref{fig1},
$\vec dl \parallel {\hat x}$ and $\vec r$ lies in the $x-y$ plane,
so that
\begin{eqnarray}
\gamma_1(t) &=& -k_0 M_z(t)/|\vec M|, \nonumber\\
k_0 &=&  \mu_0 |\vec M| l/(2 \Phi_0 a \sqrt{l^2+a^2}), \label{mpot}
\end{eqnarray}
where the geometrical factor $k_0$ can be tuned by tuning the
distance $a$ between the JJ with the nanomagnet (Fig.\ \ref{fig1}).
Moreover, in this geometry, the orbital effect of the magnetic field
do not affect the phase of the JJ since $ \vec dl \cdot \vec A_B
\sim \vec dl \cdot (\vec B \times \vec r) =0$. In this geometry, the
LL equations for the nanomagnet reads
\begin{eqnarray}
\frac{d \vec M}{dt} &=& \gamma_g (\vec M \times \vec B_{\rm eff})
\label{beffeq} \\
\vec B_{\rm eff} &=& -\frac{\delta E}{\delta \vec M} = B(M_y) \hat y
+ \frac{E_J k_0}{|\vec M|} \sin(\gamma_0(t)+\gamma_1(t))\hat z
\nonumber
\end{eqnarray}
where $B(M_y)= K M_y + B(t)$ and $\gamma_g$ is the gyromagnetic
ratio \cite{comment1}. These LL equations are to be solved along
with the constraint of constant $|\vec M|$; in what follows we shall
set $|\vec M|=M_0$. We note that Eq.\ \ref{beffeq} do not include
dissipation which shall be treated in Sec.\ \ref{gilres}. Thus the
solutions obtained in this section can be treated as limiting case
of very weakly dissipating nanomagnets. We also note that our
analysis do not take into account the change in $I_s$ arising from
the spin-flip scattering induced by the coupling of the JJ with the
nanomagnet \cite{balat1,holm1}. This can be justified by the fact
that in our geometry, the nanomagnet does not reside atop the
junction and thus we expect the spin-flip scattering matrix elements
to be small. Further, even with a significant contribution from
spin-flip scattering, such effects become important when the Larmor
frequency of the magnetization $\omega_L \ge \Delta_0/\hbar$
\cite{holm1} which is not the regime that we focus on. This issue is
discussed further in Sec.\ \ref{dissec}.

Eq.\ \ref{beffeq} represents a set of non-linear equations which, in
most cases, need to be solved numerically. Here we identify a limit
in which these equations admits an analytic, albeit perturbative,
solution for arbitrary $B(t)$. To this end we define the following
dimensionless quantities
\begin{eqnarray}
\vec m  &=& \vec M/M_0 = (\sin \theta \cos \phi, \cos \theta, \sin
\theta \sin \phi) \nonumber\\
\omega_B (t) &=& B(M_y)/B_1, \quad \epsilon_0 =  k_0 E_J/(B_1 M_0)
\nonumber\\
B(t) &=&  B_1 f(t), \quad \tau= \gamma_g B_1 t, \quad
\omega_0=\omega'_0/(\gamma_g B_1) \label{paramet1}
\end{eqnarray}
where $f(t)$ is a dimensionless function specifying the time
dependence of the magnetic field, $\omega_0$ is the dimensionless
Josephson frequency (scaled with the frequency associated with the
magnetic field $B_1$), and $B_1$ is the amplitude of the external
magnetic field. In what follows we shall seek perturbative solution
for $\vec m$ in the weak coupling and weak anisotropy limit (for
which $\epsilon_0, \, K M_0/B_1 \ll 1$ and $k_0 \le 1$) to first
order in $\epsilon_0$ and $K$. In terms of the scaled variables, the
LL equations (Eq.\ \ref{beffeq}) can be written in terms of $\theta$
and $\phi$ as
\begin{eqnarray}
\frac{d \phi}{d\tau} &=& \omega_B(\tau) - \epsilon_0 \cot \theta
\sin
\phi \sin(\gamma_0(\tau) - k_0 \sin\theta \sin\phi) \nonumber\\
\frac{d \theta}{d \tau} &=& \epsilon_0 \cos\phi \sin(\gamma_0(\tau)
- k_0 \sin\theta \sin\phi). \label{tpeq1}
\end{eqnarray}
with the initial condition $ \phi(\tau=0)=0$ and $\theta(\tau=0)=
\theta_0$. We note that the choice of this initial condition for
$\theta$ and $\phi$ amounts to choosing the initial magnetization of
the nanomagnet in the $x-y$ plane: $\vec M= (M_1, M_2,0)$ where
$\cos \theta_0 = M_2/M_0$, and $M_1^2 +M_2^2= M_0^2$. We choose
$\theta_0$ such that $\cot \theta_0 <1$ and the perturbative
solutions that we present remains valid as long as $\epsilon_0
\cot(\theta) \ll 1$. We have checked that this limit is satisfied in
all our numerical simulations described in Sec.\ \ref{numsol}.

The perturbative solutions of Eq.\ \ref{tpeq1} can be obtained by
writing
\begin{eqnarray}
\theta(\tau) &=& \delta \theta(\tau), \quad \phi(\tau)= z(\tau) +
\delta
\phi(\tau) \nonumber\\
z(\tau) &=& K \cos(\theta_0) M_0 \tau/B_1 + \int_0^{\tau} d\tau
f(\tau) \label{pertsol}
\end{eqnarray}
where $\delta \theta(\tau)$ and $\delta \phi(\tau)$ satisfies, to
first order in $\epsilon_0$ and $K$ [{\it i.e.}, neglecting terms
${\rm O}(\epsilon_0^2)$, ${\rm O}(K\epsilon_0)$ and ${\rm O}(k_0
\epsilon_0)$],
\begin{eqnarray}
\frac{d \delta \phi}{d\tau} &=& -\epsilon_0 \cot (\theta_0) \sin
(z(\tau)) \nonumber\\
&& \times \sin(\gamma_0(\tau) - k_0 \sin(\theta_0) \sin(z(\tau)))
\label{tpeq2} \\
\frac{d \delta \theta}{d \tau} &=& \epsilon_0 \cos(z(\tau))
\sin(\gamma_0(\tau) - k_0 \sin(\theta_0) \sin(z(\tau))). \nonumber
\end{eqnarray}
The solution of Eq.\ \ref{tpeq2} is straightforward and can be
written as
\begin{eqnarray}
\delta \theta(\tau) &=& \epsilon_0 \int_0^{\tau} dt'
\left[\cos(z(t'))
\sin[\gamma_0(t') \right. \nonumber\\
&& \left. - k_0 \sin(\theta_0) \sin(z(t'))] \right]\nonumber\\
\delta \phi(\tau) &=& -\epsilon_0 \cot \theta_0 \int_0^{\tau} d t'
\left[
\sin(z(t')) \sin [\gamma_0(t') \right. \nonumber\\
&& \left.- k_0 \sin(\theta_0) \sin(z(t'))]\right] \label{solgen}
\end{eqnarray}

Eqs.\ \ref{pertsol}, \ref{tpeq2}, and \ref{solgen} constitute the
central result of this work. These equations describe the dynamics
of a nanomagnet in the presence of weak coupling with a JJ. We note
that in obtaining these results, we have neglected the normal state
resistance of the JJ which can be safely done for tunnel junctions
or for weak links with large resistance and small capacitance
\cite{cai1}. We also note that the domain of validity of these
solutions require $\delta \theta(\tau), \delta \phi(\tau) \le
z(\tau)$ at all times; we shall discuss this domain in the context
of specific drives in Sec.\ \ref{pecresults}.  We now use these
solutions to study the behavior of the supercurrent of the JJ given
by
\begin{eqnarray}
I_s = I_c \sin \left[ \gamma_0(\tau) - k_0 \sin(\phi(\tau))
\sin(\theta(\tau)) \right] \label{cricur}
\end{eqnarray}
for several possible magnetic field profiles. Here $I_c$ is the
critical current of the JJ. Although Eq.\ \ref{cricur} yields $I_s$
for any magnetic field profile, in what follows we shall concentrate
on constant and periodically varying magnetic fields since they
allow for Shapiro-step like features in the I-V characteristics of a
voltage biased JJ.

Before ending this subsection, we note that the solutions for $\vec
M$ is stable against small fluctuations of the direction of the
applied magnetic field. To see this, we write the external magnetic
field $\vec B$ is applied in an arbitrary direction in the $x-y$
plane: $\vec B= B_1 f(t) ( \sin(\alpha_0), \cos(\alpha_0),0)$ with
$K \alpha_0 \ll K$. Next, we move to a rotated coordinate frame for
which the magnetization $\vec m'$ is related to $\vec m$ by
\begin{eqnarray}
\left( \begin{array} {c} m'_x \\ m'_y \\ m'_z \end{array} \right)
&=& \left( \begin{array} {ccc} \cos \alpha_0 & -\sin \alpha_0 & 0 \\
\sin \alpha_0 & \cos \alpha_0 & 0 \\ 0 & 0 & 1 \end{array} \right)
\left( \begin{array} {c} m_x \\ m_y \\ m_z \end{array} \right)
\end{eqnarray}
We proceed by using the parametrization $ \vec m' = (\sin \theta'
\cos \phi', \cos \theta', \sin \theta' \sin \phi')$. In this
representation, the initial values of $\vec m'$ are given by
\begin{eqnarray}
m'_x &=& \sin(\theta_0-\alpha_0), \,\, m'_y=
\cos(\theta_0-\alpha_0), \, m'_z=0
\end{eqnarray}
where $\theta_0$ and $\phi_0=0$ depicts the initial condition for
$\vec m$. Next, repeating the same algebraic steps as outlined
earlier in the section, one finds that the equations governing
$\theta'$ and $\phi'$ are given by
\begin{eqnarray}
\frac{d \theta^{\prime}}{d\tau} &=& \epsilon_0 \cos(\phi^{\prime})
\sin(\gamma_0(\tau)-k \sin(\theta^{\prime}) \sin(\phi^{\prime})) \\
\frac{d \phi^{\prime}}{d \tau} &=& \omega'_B(\tau)- \epsilon_0
\cot(\theta^{\prime}) \sin(\phi^{\prime}) \nonumber\\
&& \times \sin(\gamma_0 (\tau)-k_0 \sin (\theta^{\prime})
\sin(\phi^{\prime}) \nonumber\\
\omega'_B(\tau) &=& K (\cos(\alpha_0) \cos(\theta')+ \sin(\alpha_0)
\sin(\theta') \sin(\phi'))/B_1 \nonumber\\
&& + f(\tau) \simeq \omega_B(\tau) + {\rm O}(K\alpha_0)
\label{roteq}
\end{eqnarray}
Note that the analytic solution to Eq.\ \ref{roteq} can only be
obtained when terms ${\rm O}(K \alpha_0)$ can be neglected. In this
case, the perturbative solution to Eq.\ \ref{roteq} can be obtained
in the same way as done before in this section. The final result is
\begin{eqnarray}
\theta'(\tau) &=& \delta \theta'(\tau), \quad \phi'(\tau)= z(\tau) +
\delta
\phi'(\tau) \nonumber\\
\delta \theta(\tau) &=& \epsilon_0 \int_0^{\tau} dt'
\left[\cos(z(t'))
\sin[\gamma_0(t') \right. \nonumber\\
&& \left. - k_0 \sin(\theta_0 -\alpha_0) \sin(z(t'))] \right]\nonumber\\
\delta \phi(\tau) &=& -\epsilon_0 \cot \theta_0 \int_0^{\tau} dt'
\left[
\sin(z(t')) \sin [\gamma_0(t') \right. \nonumber\\
&& \left.- k_0 \sin(\theta_0 -\alpha_0) \sin(z(t'))]\right]
\label{solgen2}
\end{eqnarray}
The behavior of these solutions shall be checked against exact
numerics in Sec.\ \ref{numsol}.

\subsection{Constant and Periodically varying magnetic fields}
\label{pecresults}

In this section, we apply our perturbative results on constant and
periodically time-varying magnetic fields for which the I-V
characteristics of the JJ may have Shapiro-like steps. While this
effect has been discussed, using a somewhat different geometry, in
Ref.\ \onlinecite{cai1} for constant magnetic field, we demonstrate
its presence for periodic magnetic fields.

{\it Constant magnetic field}: This case was studied in Ref.\
\onlinecite{cai1}. For an external constant voltage, $g(t)=1$ and
one has $\gamma_0 = \omega_0 \tau +\gamma_{00}$, where $\gamma_{00}$
is the intrinsic phase difference across the JJ at $t=0$. Further,
in this case, $f(t)=1$, and $z(\tau)= \omega_c \tau$ where
$\omega_c= 1+ K M_2/B_1$. Thus the supercurrent to the leading order
and for $\epsilon_0, K \ll 1$, is given by
\begin{eqnarray}
I_s &\simeq& I_c \sin(\omega_0 \tau + \gamma_{00} -k_0 \sin(\theta_0) \sin(\omega_c \tau)) \nonumber\\
&=& I_c \sum_n J_n [k_0 \sin (\theta_0)]
\sin[(\omega_0-n\omega_c)\tau + \gamma_{00}] \label{solzero}
\end{eqnarray}
which indicates the presence of a finite DC component of $I_s$
leading to Shapiro steps in the I-V characteristics of the
JJ-nanomagnet system at
\begin{eqnarray}
\omega_0=n^0 \omega_c. \label{condcons1}
\end{eqnarray}
To study the stability of these steps we consider the solution to
${\rm O}(\epsilon_0)$. For constant magnetic field, the ${\rm
O}(\epsilon_0)$ correction to $\vec M$ can be obtained from Eq.\
\ref{solgen} which, after some straightforward algebra, yields for
$z(\tau)=\omega_c \tau$
\begin{eqnarray}
\delta \theta(\tau) &=& - \epsilon_0  \sum_{n} J_n(k_0
\sin(\theta_0)) \nonumber \\
&& \times \sum_{s=\pm1} \frac{\cos[(\omega_0
-(n-s)\omega_c)\tau+\gamma_{00}]-\cos\gamma_{00}}{\omega_0
-(n-s)\omega_c} \nonumber\\
\delta \phi(\tau) &=& \epsilon_0 \cot(\theta_0) \sum_{n} J_n(k_0
\sin(\theta_0))
\label{pertt1}\\
&& \times \sum_{s=\pm1} s \frac{\sin[(\omega_0
-(n-s)\omega_c)\tau+\gamma_{00}]-\sin \gamma_{00}}{\omega_0
-(n-s)\omega_c} \nonumber
\end{eqnarray}
We note that for $n=n_{\pm} = n^0 \mp 1$, both $\delta \theta$ and
$\delta \phi$ grows linearly in time. These terms turn out to be the
most important corrections to the zeroth order solution near
$\omega_0=n^0 \omega_c$ and leads to the destabilization of the
steps as $\epsilon_0$ increases. We also note that such terms
restrict validity of the perturbative expansion up to a finite time
$T_p$ so that $\epsilon_0 \cot(\theta_0) J_{n_{\pm}}(k
\sin(\theta_0)) T_p \sim 1$; we shall discuss this in more details
while comparing our perturbative results with exact numerics in
Sec.\ \ref{numsol}. The supercurrent to first order in $\epsilon_0$
and $K$ is thus given by
\begin{eqnarray}
I_s &\simeq& I_c \sin(\omega_0 \tau - k_0 \sin (\theta_0+ \delta
\theta(\tau))
\sin(\omega_c \tau+\delta \phi(\tau))) \label{firstorder1} \nonumber\\
\end{eqnarray}
The behavior of the DC component of $I_s$ in the presence of these
corrections is charted out in Sec.\ \ref{numsol}.

{\it Periodic Magnetic fields}: In this case, we choose a periodic
magnetic field so that $f(\tau)= \cos(\omega_1 \tau)$, where
$\omega_1$ is the external drive frequency measured in units of
$\gamma_g B_1$ \cite{commentmag}. For this choice, one has $ z(\tau)
= \omega_2 \tau + \sin(\omega_1 \tau)/\omega_1$, where $\omega_2=
\gamma_g K M_2/B_1$. Thus the zeroth order solution for the
supercurrent $I_s^{\rm periodic}$ reads
\begin{eqnarray}
I_s ^{\rm periodic} &\simeq& I_c \sin[ \omega_0 \tau + \gamma_{00} -
k_0 \sin(\theta_0) \nonumber\\
&& \times \sin ( \omega_2 \tau
+ \sin(\omega_1 \tau)/\omega_1) ] \nonumber\\
&=& I_c \sum_{n_1, n_2} J_{n_1}(k_0 \sin(\theta_0))
J_{n_2}(n_1/\omega_1) \nonumber\\
&& \times \sin[\gamma_{00} +(\omega_0 -n_2 \omega_1 - n_1 \omega_2)
\tau].
\end{eqnarray}
We note that this solution admits a finite DC component of $I_s^{\rm
periodic}$ and hence Shapiro-like steps for $(n_1,n_2)=(n_1^0,
n_2^0)$ for which
\begin{eqnarray}
\omega_0 - n_2^0 \omega_1 - n_1^0 \omega_2=0. \label{condper1}
\end{eqnarray}
The amplitude of these peaks depend on product of two Bessel
functions unlike the ones found for constant magnetic field
\cite{cai1}; moreover, the condition for their occurrence depends on
two distinct integers which allows the peaks to occur in the absence
of any DC voltage across the junction. The condition for occurrence
of such peaks are given by $\omega_0=0$ and $\omega_2= n_2^0
\omega_1/n_1^0$; they provide examples of Shapiro steps without any
voltage bias across a JJ and have no analog in their constant
magnetic field counterparts.

The first order corrections to these solutions can be obtained in a
manner analogous to one used for constant magnetic field. The final
result is given by
\begin{eqnarray}
I_s^{\rm periodic} &\simeq& I_c \sin \big[\omega_0 \tau - k_0
\sin(\theta_0 +\delta \theta_p(\tau)) \nonumber\\
&& \times \sin(\omega_2 \tau+ \sin(\omega_1 \tau)/\omega_1
+ \delta \phi_p(\tau))\big] \label{firstorder2} \nonumber\\
\end{eqnarray}
where $\delta \theta_p$ and $\delta \phi_p$ are given by
\begin{widetext}
\begin{eqnarray}
\delta \theta_p &=&
-\frac{\epsilon_0}{2}[\sum_{n_1,n_2,n_3}J_{n_1}(\frac{1}{\omega_1})J_{n_2}(k_0
\sin(\theta_0))J_{n_3}(\frac{n_2}{\omega_1})
\sum_{s=\pm1}\frac{\cos[\gamma_{00} +(\omega_0 -(n_3+s
n_1)\omega_1-(n_2+s)\omega_2)\tau]-\cos(\gamma_{00})}{\omega_0-(n_3+s
n_1)\omega_1-(n_2+s)\omega_2}] \label{percorr} \\
\delta \phi_p &=& \frac{\epsilon_0}{2} \cot(\theta_0)
[\sum_{n_1,n_2,n_3}J_{n_1}(\frac{1}{\omega_1})J_{n_2}(k_0
\sin(\theta_0))J_{n_3}(\frac{n_2}{\omega_1})\sum_{s=\pm1}
 s \frac{\sin[\gamma_{00}+(\omega_0-(n_3+s
n_1)\omega_1-(n_2+s)\omega_2)\tau]-\sin(\gamma_{00})}{\omega_0-(n_3+s
n_1)\omega_1-(n_2+s)\omega_2}] \nonumber
\end{eqnarray}
\end{widetext}
We note that the main contribution to the zeroth order results again
comes from terms linear in time which occurs for
\begin{eqnarray}
\omega_0-(n_3^s +s n_1^s)\omega_1-(n_2^s+s)\omega_2 &=& 0
\label{condpertur}
\end{eqnarray}
for $s=\pm 1$. The perturbation theory thus remain valid for $\tau
\le T_p^{'}$ so that
\begin{eqnarray}
\epsilon_0 T'_p J_{n_1^s}(\frac{1}{\omega_1})J_{n_2^s}(k_0
\sin(\theta_0))J_{n_3^s}(\frac{n_2^0}{\omega_1}) \le 1.
\label{pervalcond1}
\end{eqnarray}
The behavior of the DC component of $I_s^{\rm periodic}$ as a
function of $\epsilon_0$, as obtained from Eq.\ \ref{firstorder2} is
discussed and compared to exact numerics in Sec.\ \ref{numsol}.

{\it External AC drive}: Next, we consider the behavior of $I_s$ in
the presence of both an external magnetic field and an AC field of
amplitude $A$ and frequency $\omega_A$ so that $\gamma_0 (t)=
\gamma_{00}+ \omega_0 \tau + A \sin(\omega_A \tau)/\omega_A$. First
we consider a constant magnetic field for which $f(\tau)=1$. In this
case, using Eqs.\ \ref{pertsol} and \ref{cricur}, one obtains, to
zeroth order in $\epsilon_0$
\begin{eqnarray}
I_s &\simeq& I_c \sin \Big[\omega_0 \tau + A \sin(\omega_A
\tau)/\omega_A  \nonumber\\
&&- k_0 \sin(\theta_0) \sin \omega_c \tau +\gamma_{00} \Big]
\nonumber\\
&\simeq& I_c \sum_{n_1,n_2} J_{n_1}(A/\omega_A)
J_{n_2}(k_0\sin(\theta_0))
\nonumber\\
&& \times \sin(\gamma_{00} + (\omega_0 + n_1 \omega_A - n_2\omega_c)
\tau)
\end{eqnarray}
A finite DC component of $I_s$ leading to Shapiro like steps thus
appear in the I-V characteristics for integers $n_1^0,n_2^0$ which
satisfies
\begin{eqnarray}
\omega_0 + n_1^0 \omega_A - n_2^0 \omega_c=0. \label{condcons2}
\end{eqnarray}
The condition of occurrence of these peaks mimics those for periodic
magnetic field in the absence of external AC drive and the peak
amplitude depends on the product of two Bessel functions. We note
that the resultant Shapiro steps may occur for low AC drive
frequencies and thus could, in principle, be amenable to easier
experimental realization.

Next, we consider a periodically varying magnetic field in the
presence of external radiation. In this case, one has,
$f(\tau)=\omega_2+\cos(\omega_1 \tau)$. Using Eq.\ \ref{pertsol},
one has $z(\tau)=\omega_2 \tau + \sin(\omega_1 \tau)/\omega_1$ which
leads to (Eq.\ \ref{cricur})
\begin{eqnarray}
I_s &\simeq& I_c \sin \Big[\omega_0\tau + A \sin(\omega_A
\tau)/\omega_A - k_0 \sin(\theta_0) \times  \nonumber\\
&& \sin (\omega_2 \tau+ \sin(\omega_1 \tau))/\omega_1+ \gamma_{00} \Big] \nonumber\\
&\simeq& I_c \sum_{n_1,n_2, n_3 } J_{n_1}(A/\omega_A)
J_{n_2}(k_0\sin(\theta_0)) J_{n_3}(n_2/\omega_2) \nonumber\ \\
&& \times \sin(\gamma_{00} + (\omega_0 + n_1 \omega_A - n_2 \omega_2
- n_3 \omega_1) \tau)
\end{eqnarray}
Thus the presence of the steps now occurs for a set of integers
$(n_1^0,n_2^0,n_3^0)$ which satisfies
\begin{eqnarray}
\omega_0 + n_1^0 \omega_A - n_2^0 \omega_2 -n_3^0 \omega_1=0.
\label{condper2}
\end{eqnarray}
The perturbative $O(\epsilon_0)$ corrections to the above solutions
can be carried out in similar manner to that outlined above.

\subsection{Dissipative nanomagnets}
\label{gilres}

In this section, we include the dissipative (Gilbert) term in the
LLG equations to model dissipative nanomagnets and seek a solution
to these equations in the limit weak dissipation and weak coupling
between the JJ and the nanomagnets. The resultant LLG equations are
given by
\begin{eqnarray}
\frac {d \vec M}{dt} &=& \gamma_g  (\vec M \times \vec B_{\rm
eff})-\frac{\eta \gamma_g}{M_0} \vec M \times d \vec M/dt
\label{llg1}
\end{eqnarray}
where $\eta$ is a dimensionless constant specifying the strength of
the dissipative term. Following the same parametrization as in Eqs.\
\ref{paramet1}, we find that one can express the LLG equations, in
terms of $\theta$ and $\phi$, as
\begin{eqnarray}
\frac{d \phi}{d\tau} &=& \omega_B(\tau) - \epsilon_0 \cot \theta
\sin
\phi \sin(\gamma_0(\tau) - k_0 \sin(\theta) \sin(\phi)) \nonumber\\
\frac{d \theta}{d \tau} &=& \epsilon_0 \cos(\phi)
\sin(\gamma_0(\tau) -
k_0 \sin(\theta) \sin(\phi)) \nonumber\\
&& - \eta  \omega_B(\tau) \sin(\theta). \label{tpgil1}
\end{eqnarray}
where we have neglected terms $O(\epsilon_0 \eta)$ and ${\rm
O}(\eta^2)$. We note that the effect of the dissipative term
manifests itself in $\theta$ but not in $\phi$; this fact can be
understood as a consequence of the fact that to leading order $\vec
M \times (\vec M \times \vec B_{\rm eff})$ lies along $\hat y$ and
hence only effects the dynamics of $M_y$ which depend only on
$\theta$. For small $\epsilon_0$ and $\eta$, Eq.\ \ref{tpgil1}
therefore admits a perturbative solution
\begin{eqnarray}
\phi(\tau) &=& z(\tau) + \delta \phi(\tau), \quad \theta(\tau)=
\delta \theta_d(\tau)
\nonumber\\
\delta \theta_d (\tau) &=& 2 \arctan[\tan(\theta_0/2)e^{-\eta
z(\tau)}] +\delta \theta(\tau). \label{thetadsol}
\end{eqnarray}
where $z(\tau)$, $\delta \theta(\tau)$, and $\delta  \phi(\tau)$ are
given by Eq.\ \ref{pertsol} and we have neglected terms ${\rm
O}(\epsilon_0 \eta)$. The supercurrent, in the presence of the
dissipative term is given by
\begin{eqnarray}
I_s &=& I_c \sin [\gamma_0(\tau)- k_0 \sin(\theta_d(\tau))
\sin(\phi(\tau))] \label{supcurrent}
\end{eqnarray}
The fate of the DC component of $I_s$ leading to Shapiro-like steps
in the presence of the dissipative term shall be checked numerically
in Sec.\ \ref{numsol} for both periodic and constant applied
magnetic fields.

\section{Numerical results}
\label{numsol}

In this section, we analyze the coupled JJ-nanomagnet system both
without and in the presence of dissipation and compare these
results, wherever applicable, to the theoretical results obtained in
Sec.\ \ref{genres}. In what follows, we focus on cases of constant
or periodically varying magnetic fields since Shapiro-step like
features are expected to appear in the I-V characteristics of the JJ
only for these protocols. The LLG equations for magnetization solved
numerically to generate the data for the plots are given by
\begin{eqnarray}
\frac{d m_x}{d \tau}&=&[-\beta_1 (1+\eta ^2 m_x^2)-\beta_2 \eta
(m_z+ \eta  m_x m_y) \nonumber\\
&& + \beta_3 \eta (m_y- \eta m_x
m_z)]/(1+\eta ^2) \nonumber \\
\frac{d m_y}{d \tau}&=&[-\beta_2 (1+\eta ^2 m_y^2)-\beta_3 \eta
(m_x+ \eta  m_z m_y) \nonumber\\
&& + \beta_1 \eta (m_z- \eta m_x m_y)]/(1+\eta ^2)
\nonumber\\
\frac{d m_z}{d \tau}&=&[-\beta_3 (1+\eta ^2 m_z^2)-\beta_1 \eta
(m_y+ \eta  m_z m_x) \nonumber\\
&& + \beta_2 \eta (m_x- \eta m_z m_y)]/(1+\eta ^2)
 \label{cartnum}
\end{eqnarray}
where
\begin{eqnarray}
\beta_1 &=&-(f(\tau) \cos \alpha_0 +K' m_y) m_z \nonumber\\
&& +\epsilon_0 m_y \sin(\gamma_0(\tau)-k_0 m_z) \label{betadef} \\
\beta_2&=& -\epsilon_0 m_x \sin(\gamma_0(\tau)-k_0 m_z) +
\sin(\alpha_0) f(\tau) m_z  \nonumber\\ \ \beta_3&=& (f(\tau)
\cos(\alpha_0) +K' m_y) m_x -\sin(\alpha_0) f(\tau) m_y \nonumber
\end{eqnarray}
In these equations $f(\tau)=1$ for constant and
$f(\tau)=\cos(\omega_1 \tau)$ for the periodically varying magnetic
fields, $\alpha_0=0$ indicates an applied magnetic field along $\hat
y$, we have set $\theta_0=\pi/3$ and $\gamma_{00}=\pi/2$ for all
simulations, and $K'= KM_0/B_1$. Note that Eq.\ \ref{cartnum}
reduces to the usual LL equations for $\eta=0$. The supercurrent is
then computed using the values of $m_{z}$ obtained from Eq.\
\ref{cartnum}: $I_s = I_c \sin(\gamma_0(\tau)- k_0 m_z)$.

\begin{figure}[ht]
\centering
\includegraphics[width=4.2cm,height=3.2cm]{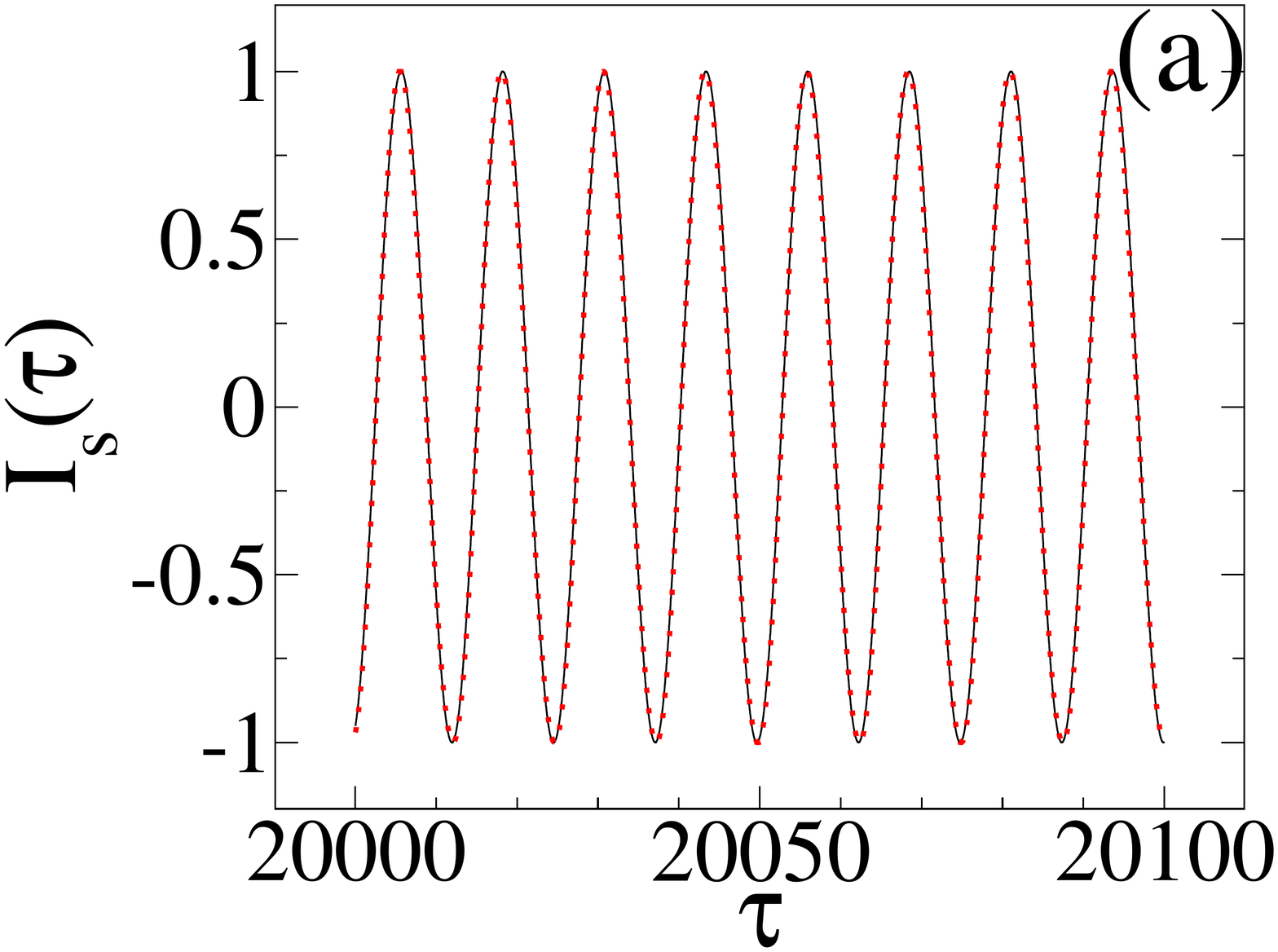}
\includegraphics[width=4.2cm,height=3.2cm]{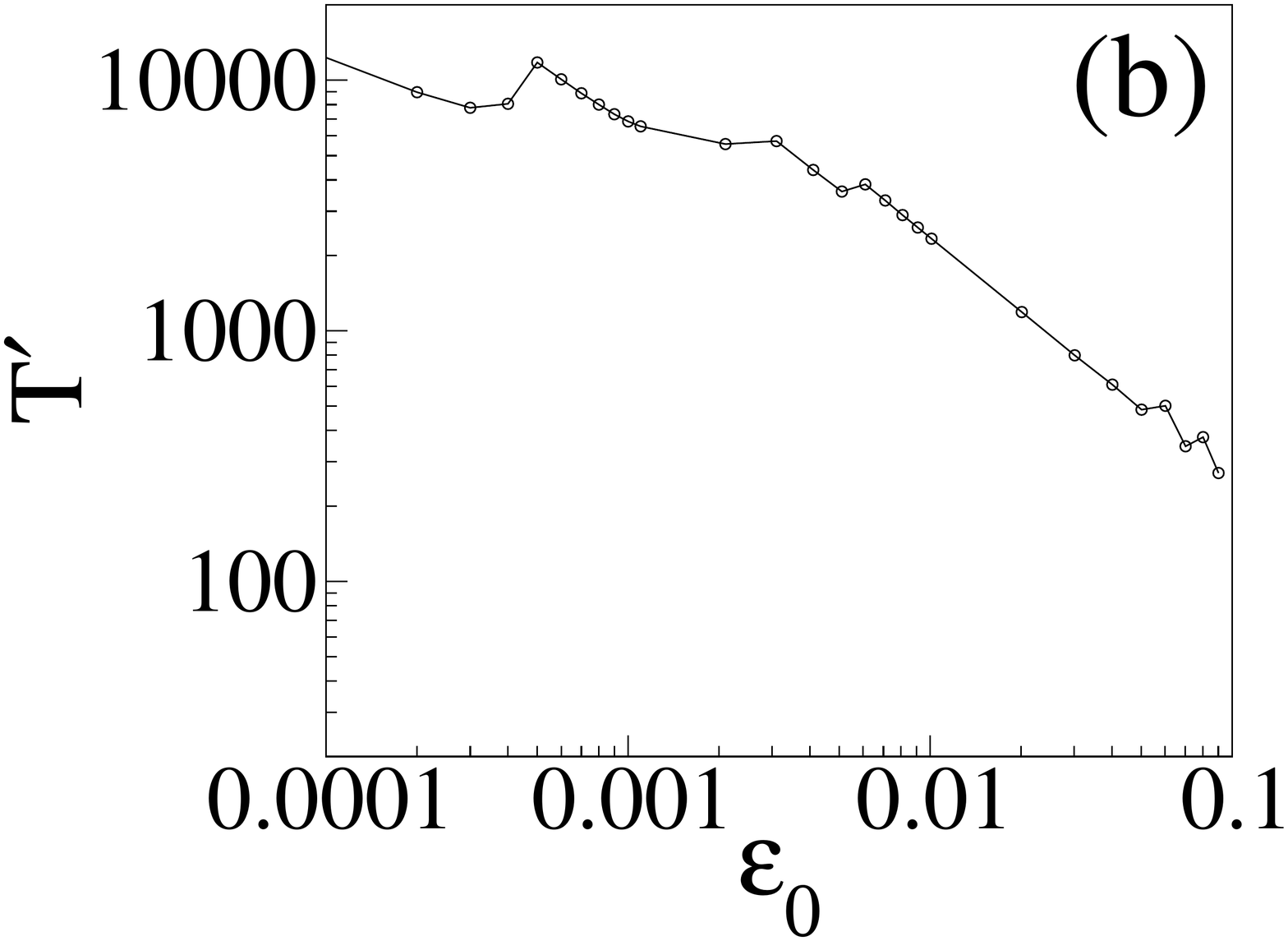}
 \caption{(a) Comparison between theoretical (red dots)
and numerical (black solid line) values of $I_s(\tau)/I_c$ as a
function of time at late times $\tau \ge 2\times 10^4$ for a
constant magnetic field $\omega_B=0.5$ along $\hat y$. Other
parameters are $\epsilon_0=10^{-4}$, $\eta=0$, $k_0=0.05$,
$K=0.0001$ and $\omega_0=0.5$. (b) Plot of time $T'$ after which the
theoretical and analytic results for $I_s(t)$ deviates by more that
$1\%$ at the peak position as a function of $\epsilon_0$.}
\label{fig2}
\end{figure}

\begin{figure}[ht]
\centering
\includegraphics[width=4.2cm,height=3.2cm]{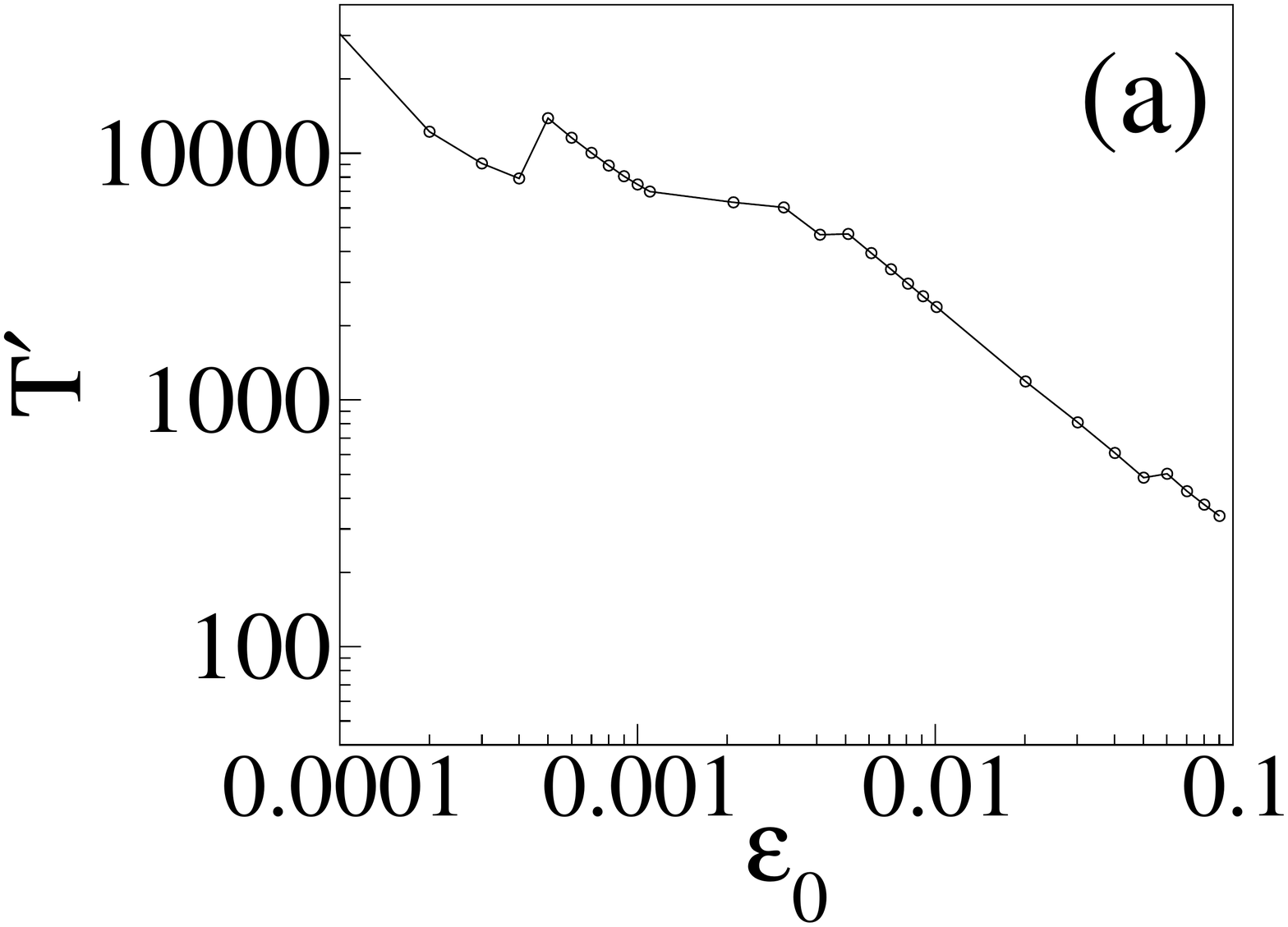}
\includegraphics[width=4.2cm,height=3.2cm]{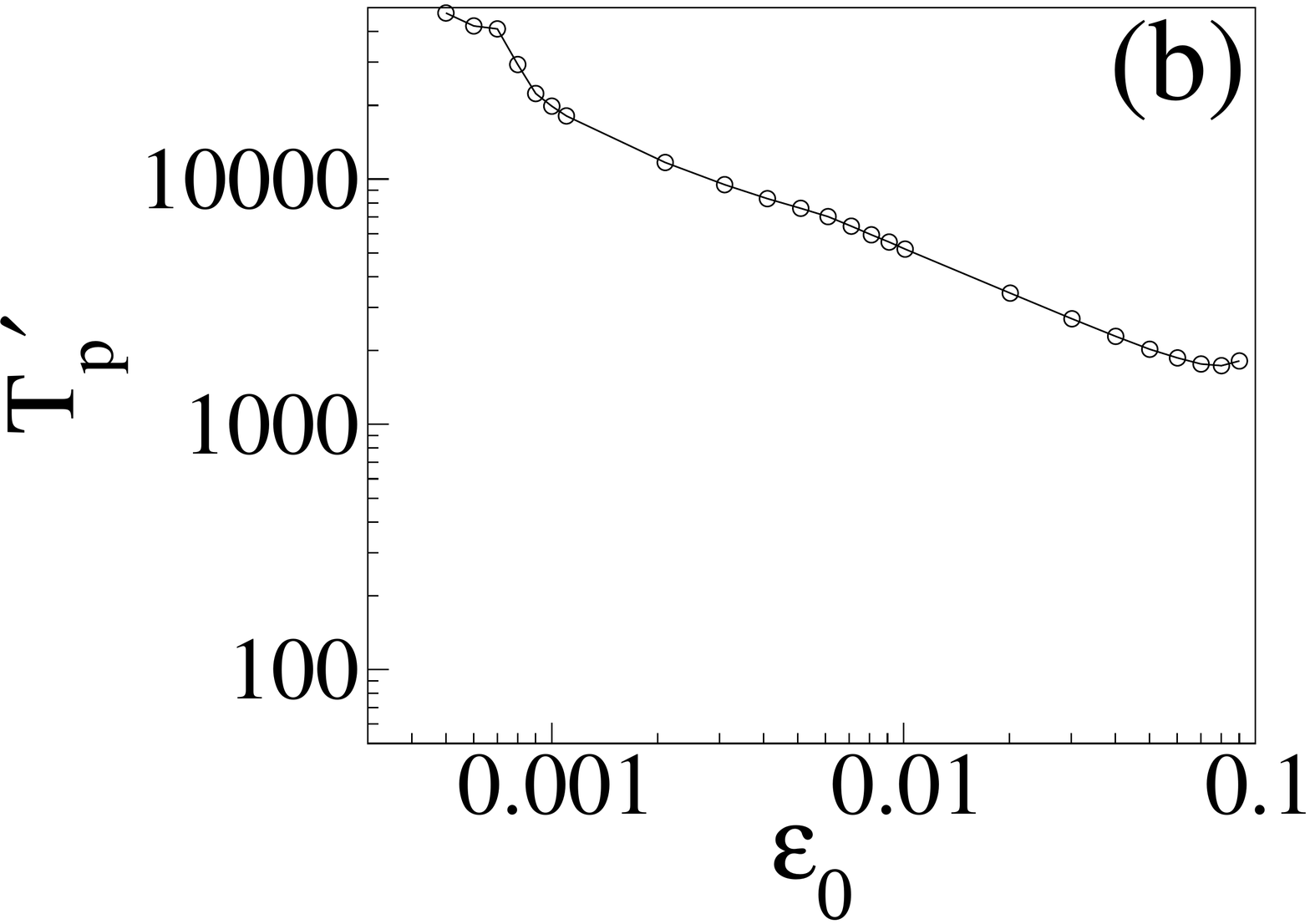}
 \caption{(a) Plot of $T'$ as a function of $\epsilon_0$
for $\eta=0.0001$. (b) Plot of $T'$ as a function of $\epsilon_0$ for
periodic protocol with $\omega_B=0$. $\eta=0$, and $\omega_1=1.2$.
All other parameters are same as in Fig.\ \ref{fig2}.} \label{fig3}
\end{figure}

To compare the theoretical results with exact numerics, we first
compare the values $I_s(\tau)/I_c$ computed theoretically (Eq.\
\ref{firstorder1}) with exact numerical result. For comparing the
two results, we have fixed the external voltage $\omega_0= \omega_B$
which leads to a Shapiro step in the I-V characteristics of the JJ
with $n^0=1$. As discussed in Sec.\ \ref{pecresults}, one expects
one of the perturbative terms to grow linearly in time in this case;
the presence of this linear term is expected to invalidate the
perturbative theoretical results for $\tau >T' \sim
\epsilon_0^{-1}$. In Fig.\ \ref{fig2}(a), we show the comparison
between theoretical and numerical values of $I_s(\tau)/I_c$ at
$\tau>2\times 10^4$ for $\epsilon_0= 10^{-4}$; we find that the
numerical and analytical values differ by less than $5\%$ even at
late times ($t \simeq 2 T'$). In Fig.\ \ref{fig2}(b), we plot $T'$,
which is the minimum time at which the deviation between theoretical
and numerical values of $I_s(\tau)/I_c$ reaches $1\%$ near the peak
position, as a function of $\epsilon_0$; the result shows the
expected decrease of $T' \sim 1/\epsilon_0$ as $\epsilon_0$
increases. In Fig.\ \ref{fig3}(a), we carry out a similar comparison
for dissipative nanomagnets with $\eta=0.0001$; we find that $T'$
decreases with $\epsilon_0$ in a qualitatively similar manner to the
non-dissipative case. However, we note that the value of $T'$ with
finite $\eta$ (Fig.\ \ref{fig3}(a)) is larger than its $\eta=0$
counterpart (Fig.\ \ref{fig2}(b)); this feature is a consequence of
opposite signs of the correction terms due to $\epsilon_0$ (Eq.\
\ref{pertt1}) and $\eta$ (Eq.\ \ref{thetadsol}). For small $\eta$
and $\epsilon_0$, these corrections tend to mutually cancel leading
to a better stability of the zeroth order result which results in
higher value of $T'$. Finally, in Fig.\ \ref{fig3}(b), we plot
$T'_p$ for a periodically varying magnetic field with
$\omega_1=1.2$. As expected from Eq.\ \ref{pervalcond1}, $T'_p \sim
100 \epsilon_0 ^{-1} \gg \epsilon_0^{-1}$ which implies much better
stability for the Shapiro-like steps for periodic magnetic field
compared to their constant field counterparts.

\begin{figure}[ht]
\centering
\includegraphics[width=4.2cm,height=3.2cm]{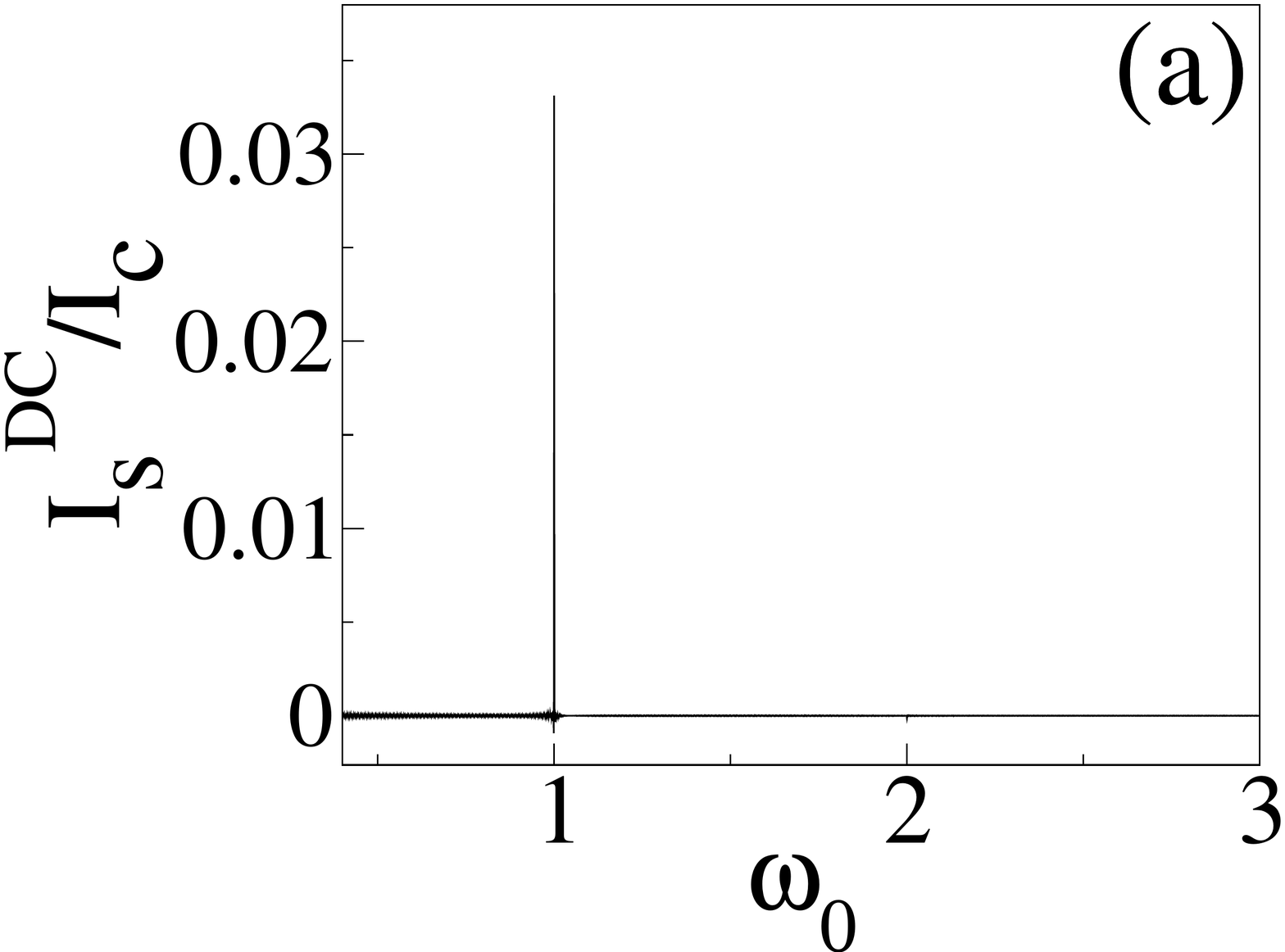}
\includegraphics[width=4.2cm,height=3.2cm]{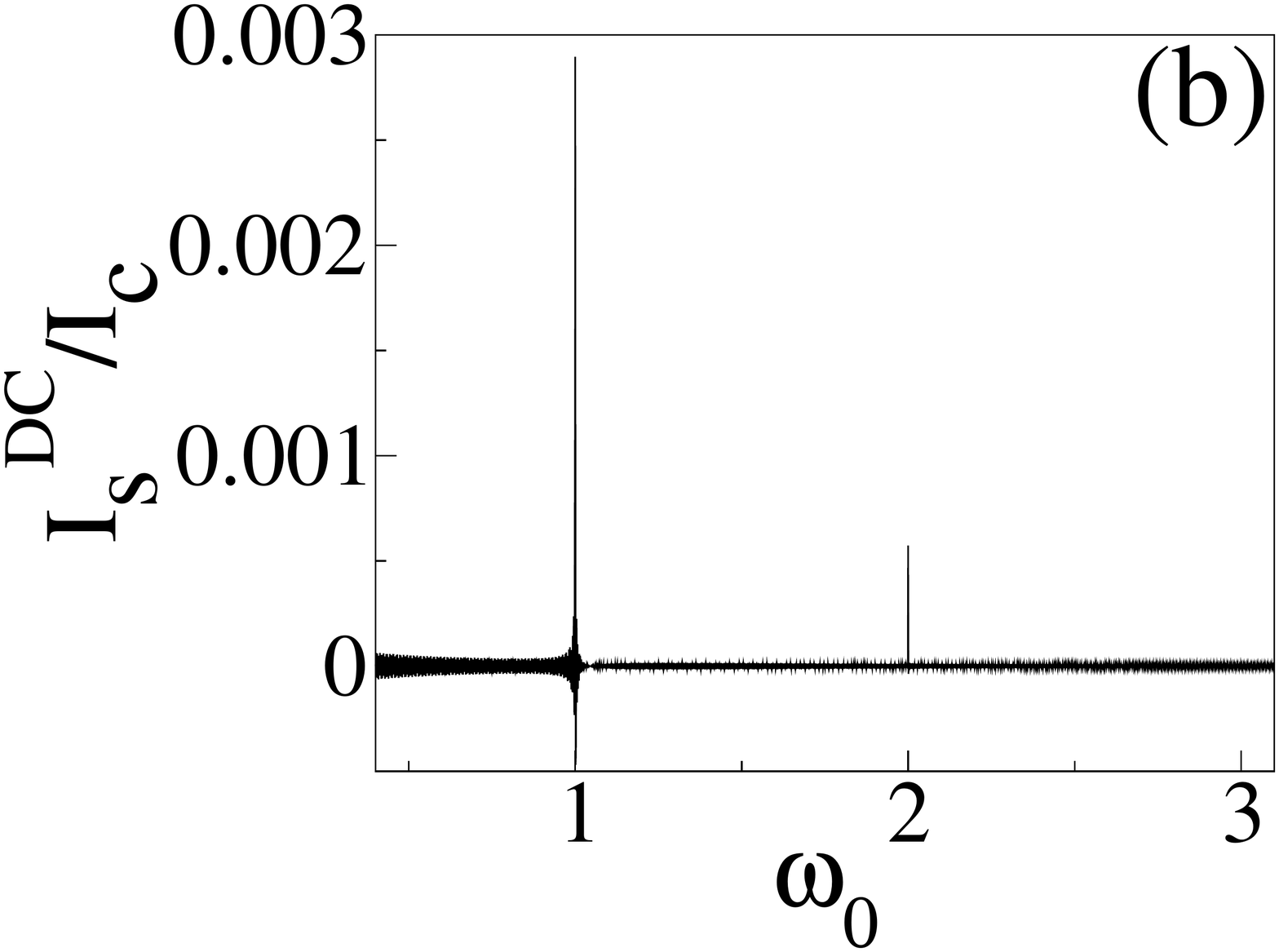}
\includegraphics[width=4.2cm,height=3.2cm]{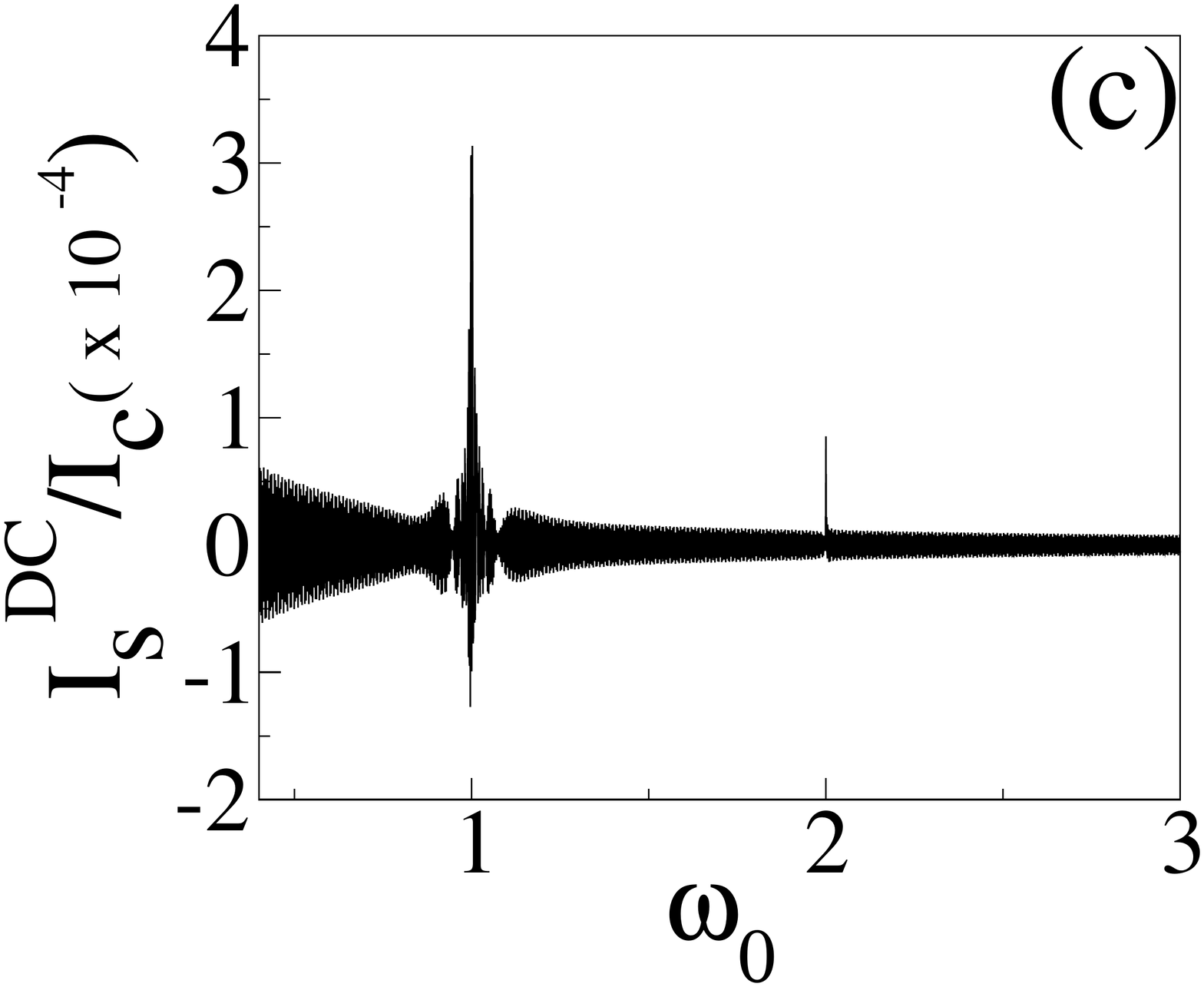}
\includegraphics[width=4.2cm,height=3.2cm]{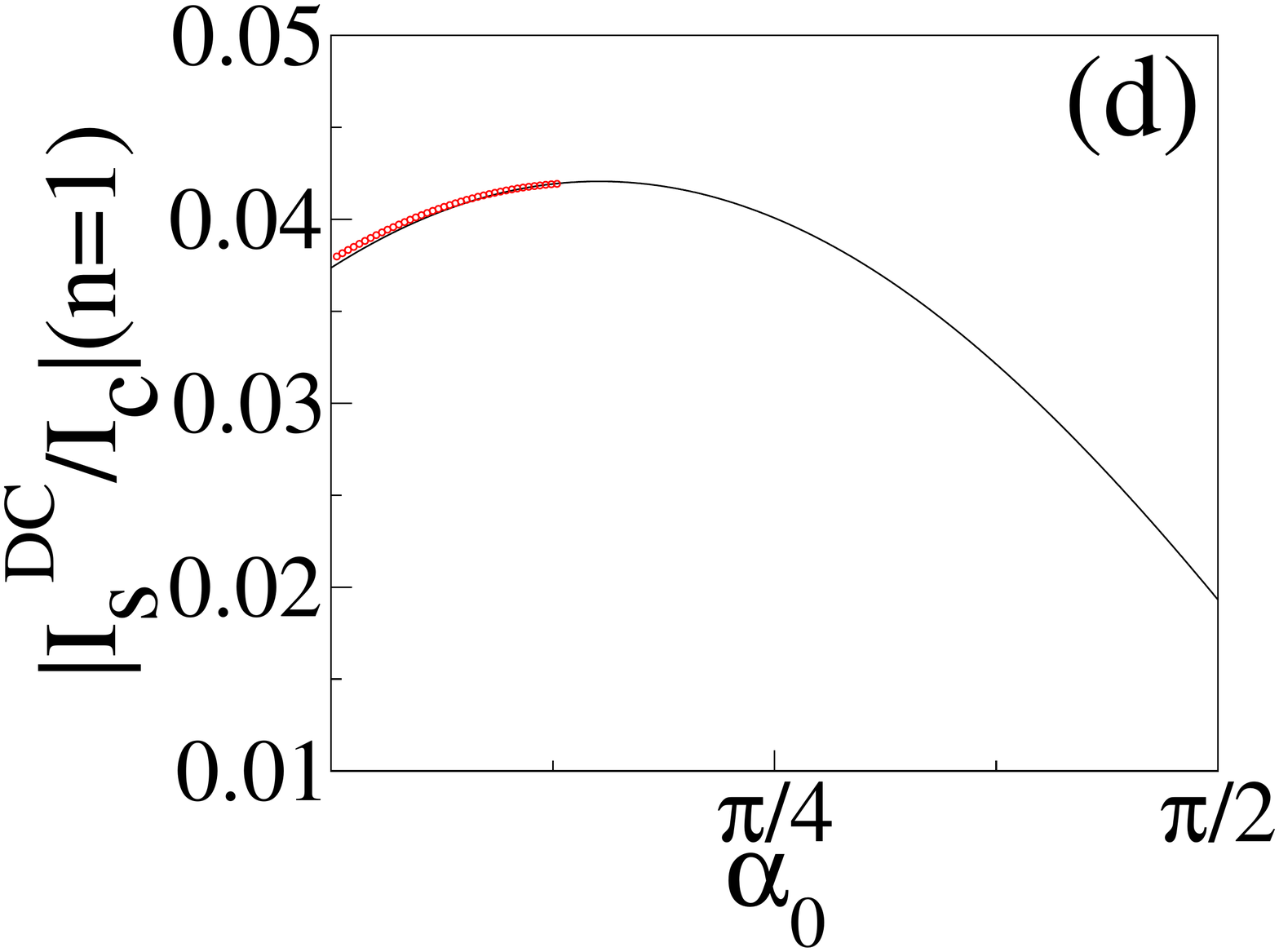}
\caption{ Plot $I_s(\omega=0)/I_c \equiv I_s^{\rm DC}/I_c$ as a
function of the Josephson frequency $\omega_0= 2e V_0/(\hbar
\gamma_gB_1)$ for a constant magnetic field $\omega_B \simeq 1$ with
$K=0.0001$, $k_0=0.1$ and (a)$\epsilon_0=0.0001$
(b)$\epsilon_0=0.001$ and (c) $\epsilon_0=0.01$. The position of the
peaks corresponds to $n^0=1$ and $n^0=2$ as predicted by theoretical
analysis. (d) Plot of the peak height for the $n^0=1$ peak as a
function of the angle $\alpha_0$ made by $\vec B$ with $\hat y$ for
$\epsilon_0=10 K =0.0001$ and $k_0=0.1$. The red dots correspond to
results from perturbative theoretical analysis near $\alpha_0=0$. }
\label{fig4}
\end{figure}

Next, we study the presence of a finite DC component of $I_s$ in the
case of a constant applied magnetic field along $\hat y$
($f(\tau)=1$ and $\alpha_0=0$) in the absence of dissipation
($\eta=0$) and external AC voltage ($\gamma_0(\tau)= \omega_0
\tau$). The results of our study is shown in Fig.\ \ref{fig4} where
we plot $I_s^{\rm DC}/I_c$, with $I_s^{\rm DC}$ given by
\begin{eqnarray}
I_s^{\rm DC} = \frac{1}{T_{\rm max}} \int_0^{T_{\rm max}} I_s(t')
dt' = I_s(\omega=0),
\end{eqnarray}
as a function of $\omega_0$ for a fixed constant $\omega_B$. Here
$T_{\rm max} = 40,000$ represents the maximum time up to which we
average $I_s(\tau)$. Note that $I_s(\tau)$ is chosen so that
increasing it any further does not lead to a change in the peak
height for $\epsilon_0=0$. As shown in Fig.\ \ref{fig4}(a), (b) and
(c), we find that for $\epsilon_0 \ll 1$, $I_s^{\rm DC}$ shows sharp
peaks at $\omega_0 = \omega_B, 2 \omega_B$ corresponding to
$n^0=1,2$ in Eq.\ \ref{condcons1}; the position of this peaks match
exactly with our theoretical results. However, the peak heights turn
out to be smaller than that predicted by theory and they rapidly
decrease with increasing $\epsilon_0$. This mismatch between
theoretical and numerical results is a consequence of the linearly
growing perturbative terms $\sim \epsilon_0$ in expression for
$\delta \theta(\tau)$ and $\delta \phi(\tau)$ (Eq.\ \ref{pertt1})
which invalidate the theoretical result for $T' \sim
\epsilon_0^{-1}$. Thus for constant magnetic field and moderate
$\epsilon_0 >0.01$, the step-like feature predicted in Eq.\
\ref{condcons1} disappears. In Fig.\ \ref{fig4}(d), we study the
behavior of the peak with variation of  $\alpha_0$. We find that the
height of the peak increases with $\alpha_0$ for small $\alpha_0$ in
accordance with the theoretical prediction of Sec.\ \ref{genres}.
For larger $\alpha_0>\alpha_0^{\rm max}$, the peak height starts to
decrease and the peak height becomes almost half of its maximum for
$\alpha_0= \pi/2$ when $\vec B
\parallel \hat x$.

\begin{figure}[ht]
\centering
\includegraphics[width=4.2cm,height=3.2cm]{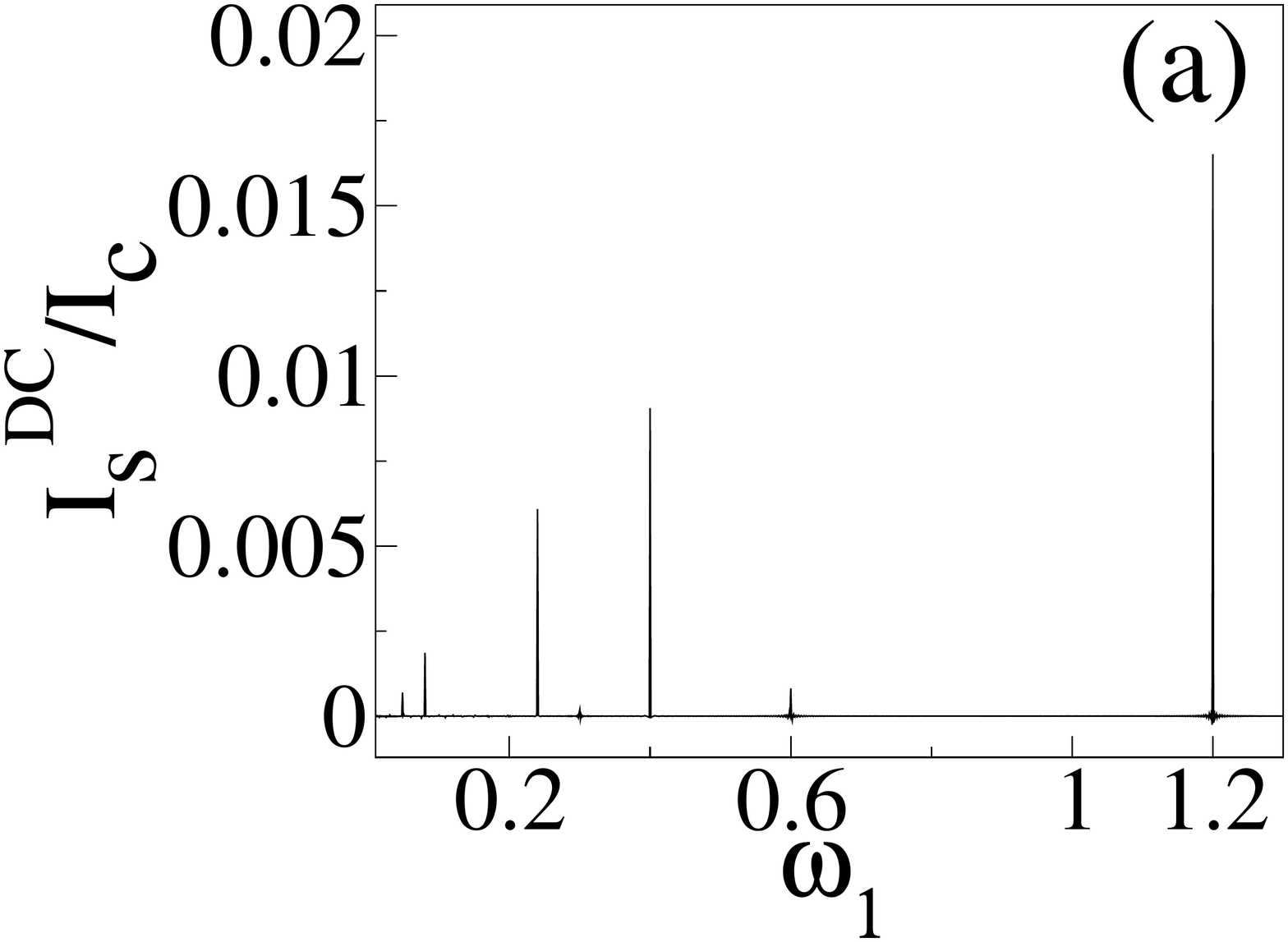}
\includegraphics[width=4.2cm,height=3.2cm]{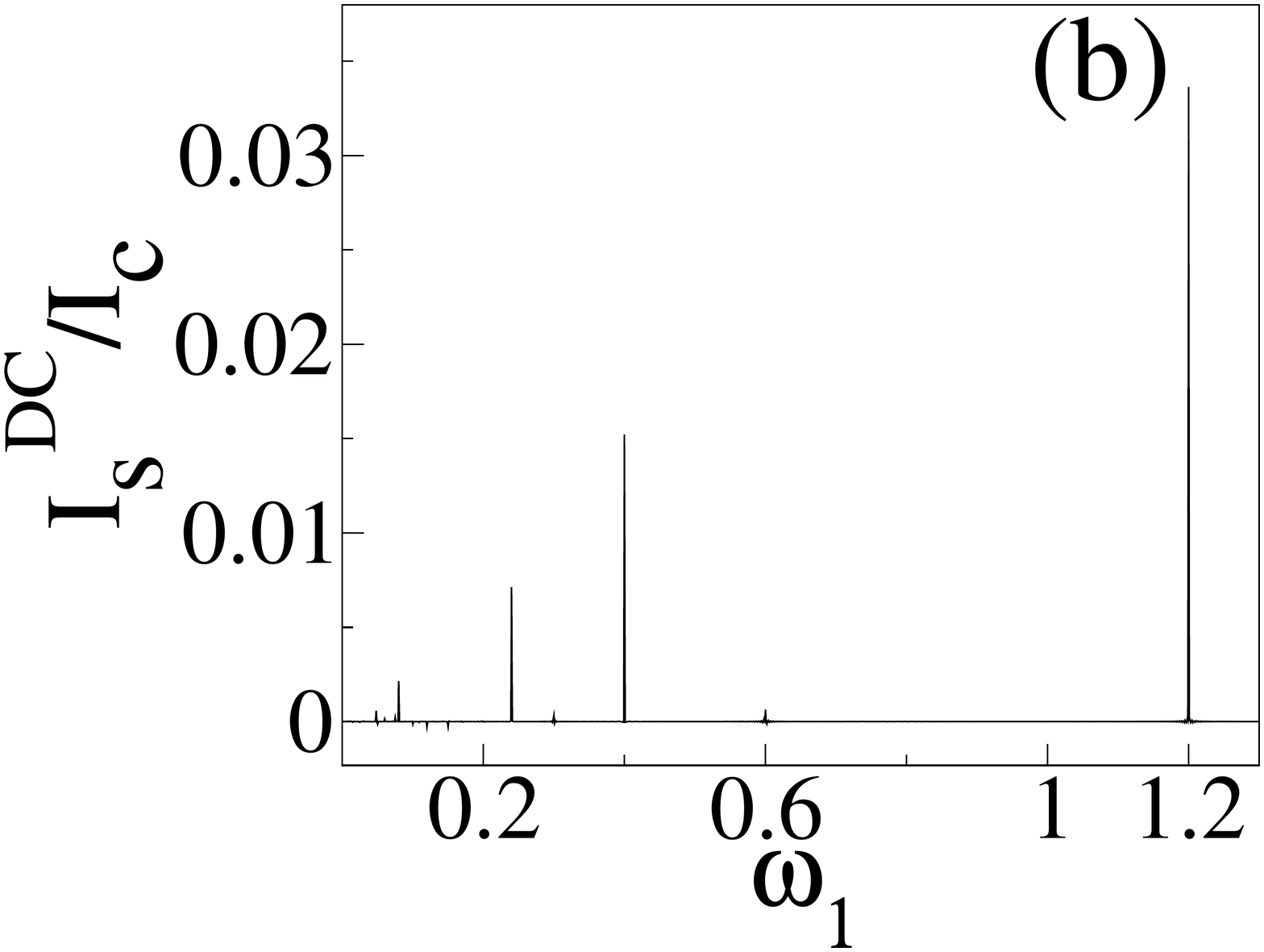}
\includegraphics[width=4.2cm,height=3.2cm]{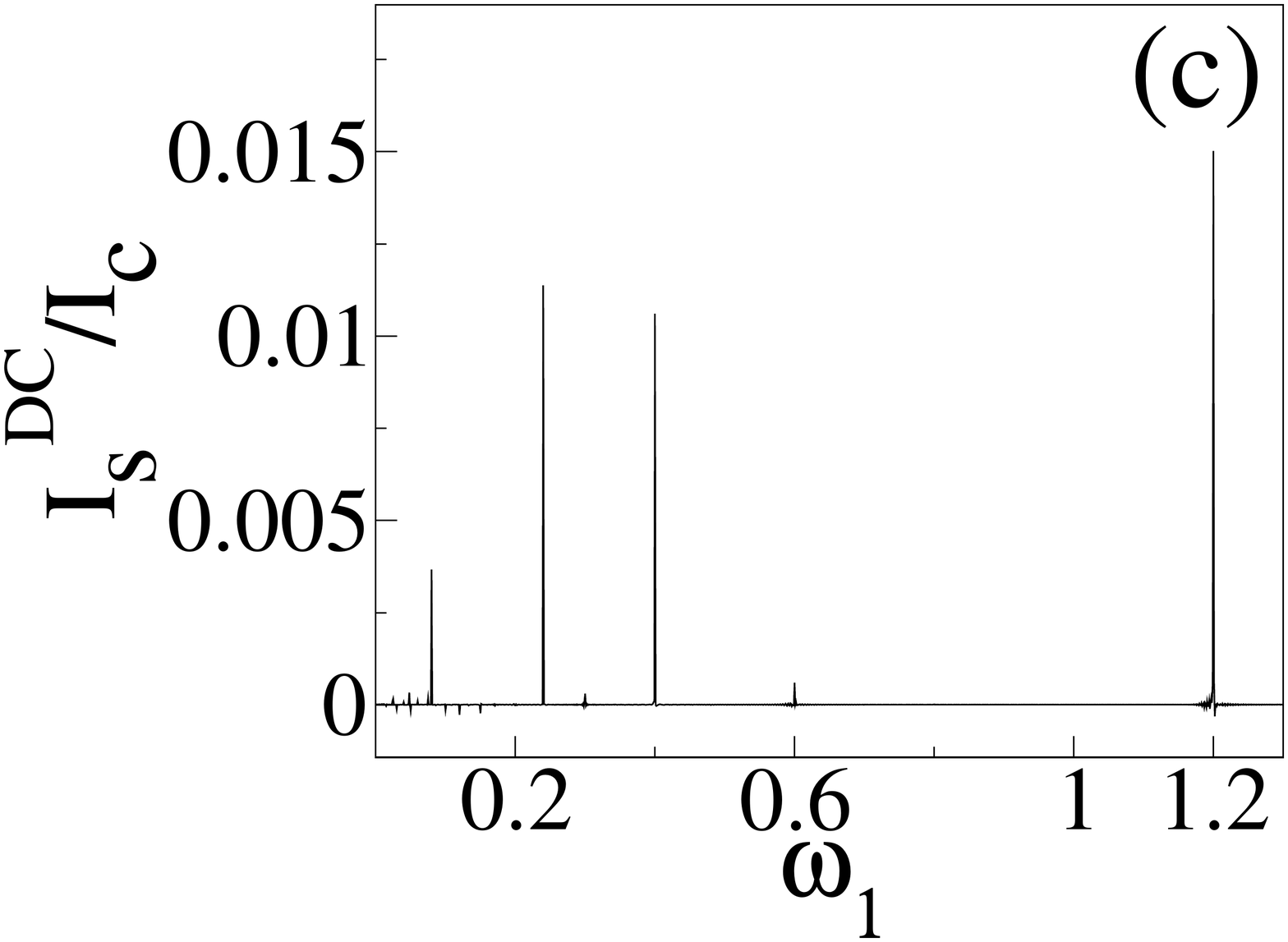}
\includegraphics[width=4.2cm,height=3.2cm]{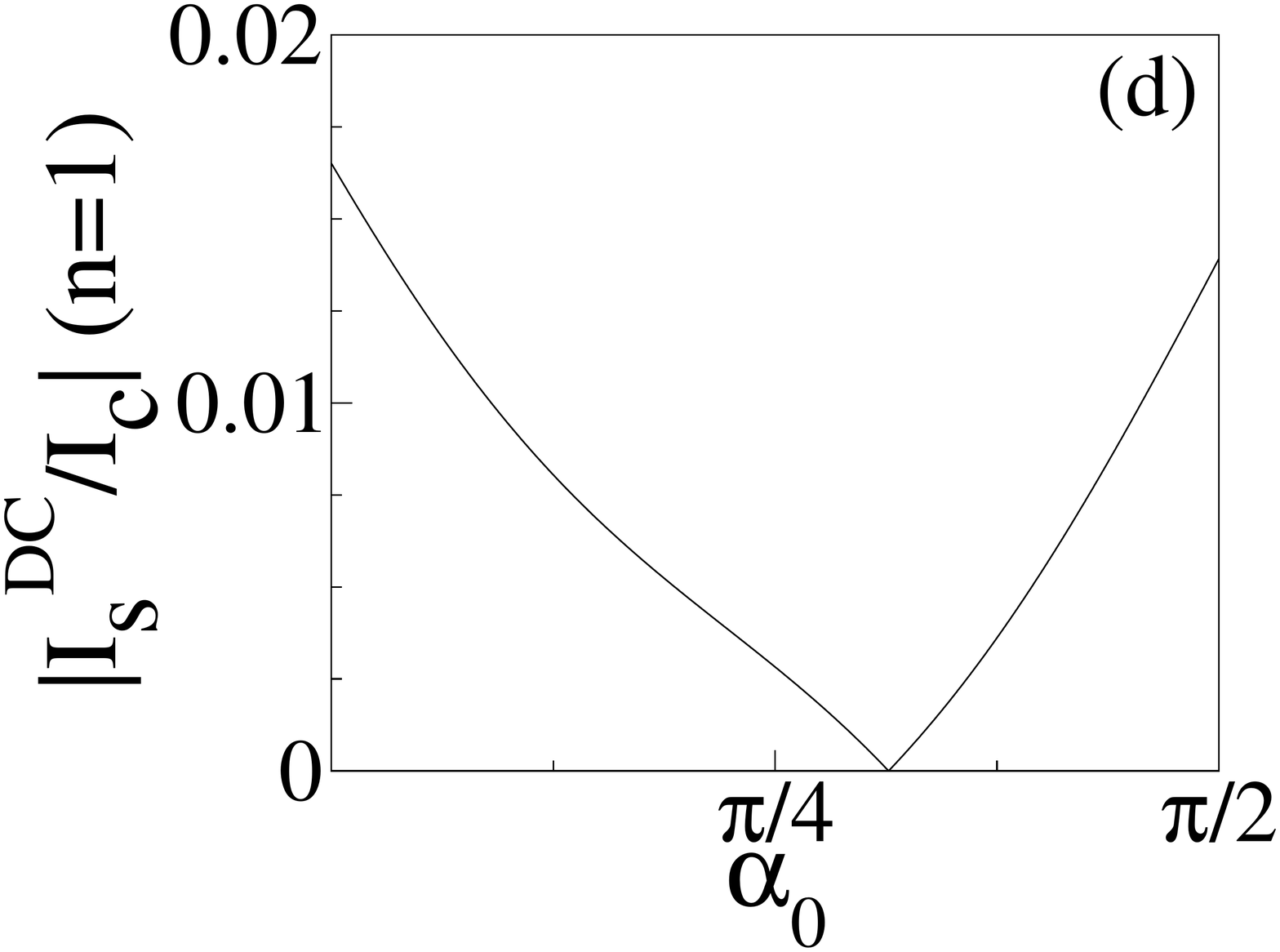}
 \caption{Plot $I_s^{\rm DC}/I_c$ for a periodically varying
magnetic field $B= B_1 \sin \omega_1 t$ as a function of $\omega_1$
with $K=0.0001$, $k_0=0.1$, $\omega_0=1.2$ and
(a)$\epsilon_0=0.0001$ (b)$\epsilon_0=0.001$ and (c)
$\epsilon_0=0.01$. The position of the peaks corresponds to
$n_2^0=1,2,3,4$ (from right to left) as predicted by theoretical
analysis. (d) Plot of the peak height for the $n_2^0=1$ peak as a
function of the angle $\alpha_0$ made by $\vec B$ with $\hat y$ for
$\epsilon_0=K =0.0001$ and $k_0=0.1$.} \label{fig5}
\end{figure}

Next, we study the characteristics of the peaks in $I^{\rm DC}_s$
for periodically varying magnetic field for which $f(\tau)=
\cos(\omega_1 \tau)$. In Figs.\ \ref{fig5}(a), (b), and (c), we plot
$I^{\rm DC}/I_c$ a function of $\omega_1$ for a fixed $\omega_0=
1.2$, $\alpha_0=\omega_2=\eta=0$, and for several values of
$\epsilon_0$. We find that the position of the peaks corresponds to
integer values of $n_2^0$ (as indicated in the caption of Fig.\
\ref{fig5}) in complete accordance with Eq.\ \ref{condper1} with
$\omega_2=0$. Moreover, in contrast to the constant magnetic field
case, the peaks of $I^{\rm DC}_s$ are much more stable against
increasing $\epsilon_0$. This features of the peaks can be
understood as follows. For periodic magnetic field with
$\omega_2=0$, the zeroth order solution is given by $z(\tau) =
\sin(\omega_1 \tau)/\omega_1$; thus the perturbative terms $\delta
\theta(\tau)$ and $\delta \phi(\tau)$ (Eq.\ \ref{percorr}) involve
product of Bessel functions. This renders the effective perturbative
parameter to be $\epsilon_0^{\rm eff} \simeq \epsilon_0
J_{n_1^s}(\frac{1}{\omega_1})J_{n_2^s}(k_0
\sin(\theta_0))J_{n_3^s}(\frac{n_2^0}{\omega_1})$ (Eqs,\
\ref{condpertur} and \ref{pervalcond1}). Consequently, the effect of
the perturbative correction to the weak coupling solution is
drastically reduced in this case leading to a better stability of
peak height with increasing $\epsilon_0$. Thus periodic magnetic
fields are expected to lead to enhanced stability of Shapiro steps
compared to their constant field counterparts. Finally in Fig.\
\ref{fig5} (d), we show the variation of the peak height of $I^{\rm
DC}_s$ as a
\begin{figure}[ht]
\centering
\includegraphics[width=4.2cm,height=3.2cm]{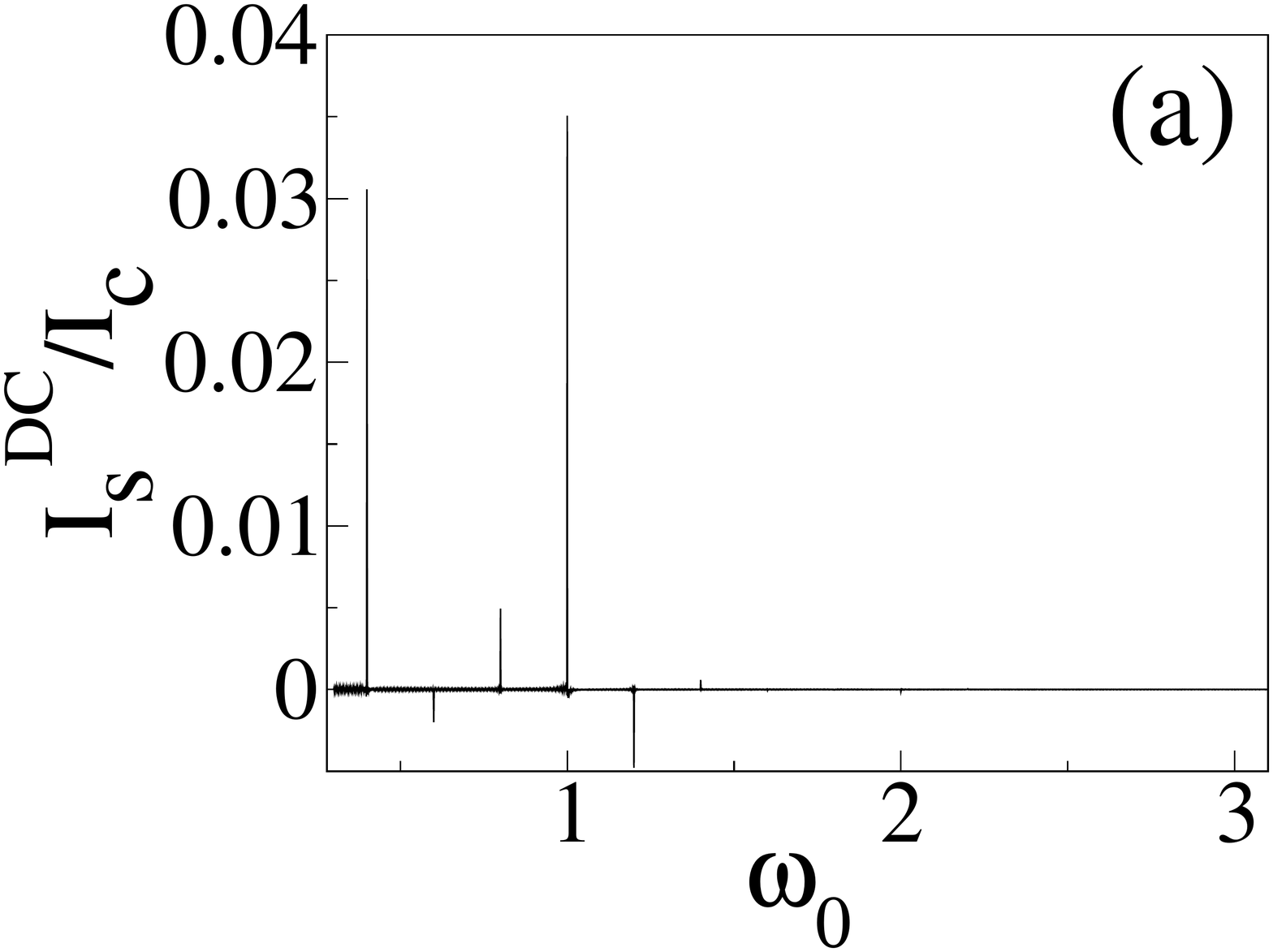}
\includegraphics[width=4.2cm,height=3.2cm]{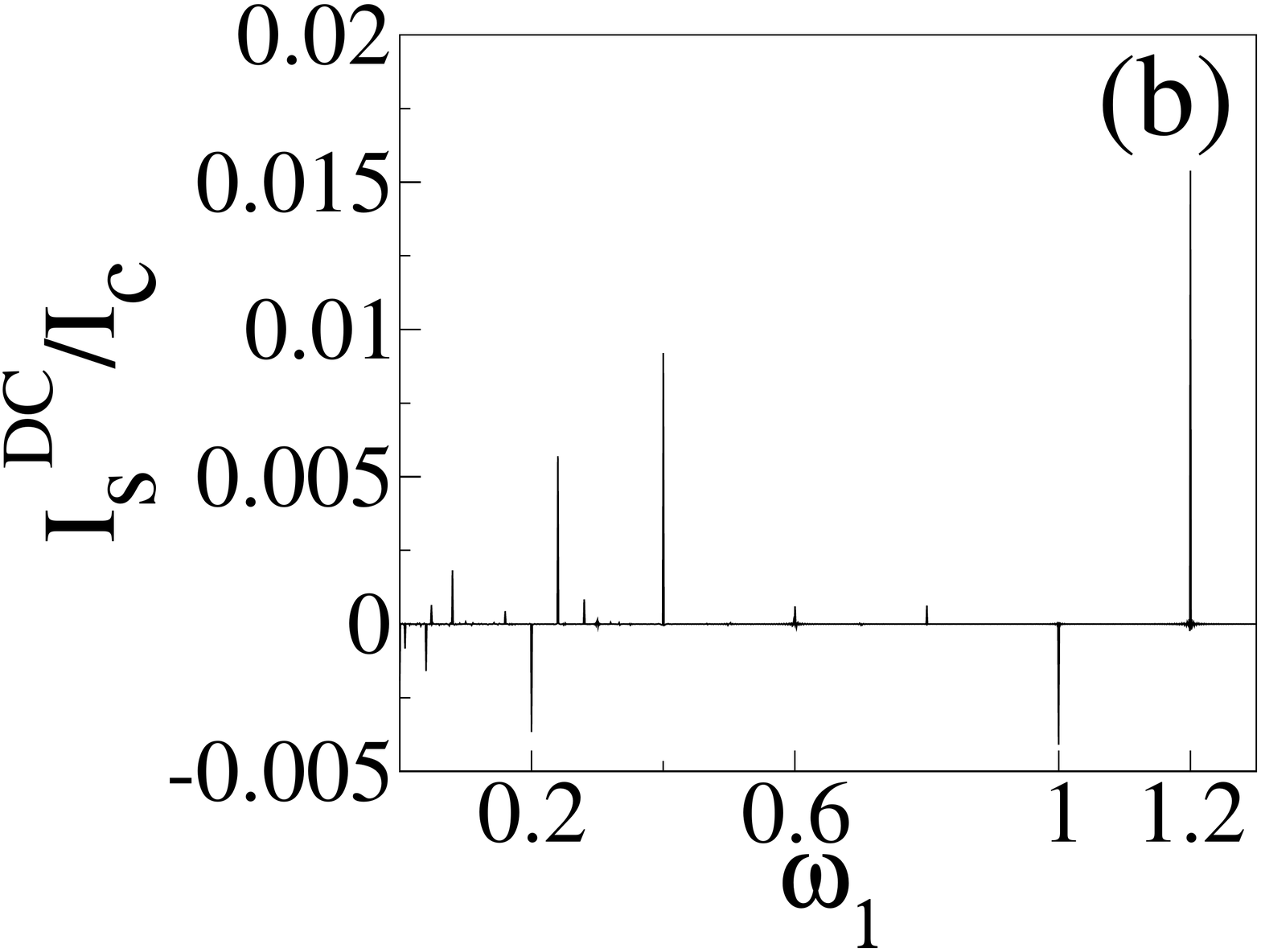}
\caption{(a) Plot of $I_s^{\rm DC}/I_c$ in the presence of an AC
field $\omega(\tau)=\omega_0 + A \sin(\omega_A \tau)/\omega_A$ with
$A=0.1$, $\omega_A=0.2$, $\epsilon_0=K=0.0001$, $k_0=0.1$ as a
function of $\omega_0$ for constant magnetic field $\omega_B=1$. (b)
Similar plot as a function of $\omega_1$ for periodic magnetic field
with $\omega_0=1.2$. All the peak positions conform to the
theoretical prediction in Sec.\ \ref{pecresults}.} \label{fig6}
\end{figure}
function of $\alpha_0$. We again find similar non-monotonic behavior
of the peak height as a function of $\alpha_0$; the reason for this
is similar to that already discussed in the context of constant
magnetic field case. However, in the present case, the correction
terms are much smaller and the peak height is accurately predicted
by the zeroth order perturbative results: $ I^{\rm DC}_s/I_c \sim
2J_{n_1^0}(k_0\sin(\theta_0 -\alpha_0)) J_{n_2^0}(n_1^0/\omega_1)$.
This is most easily checked by noting that the peak height vanishes
for $\alpha_0= \theta_0= \pi/3$ for which $ J_{n_1^0}(0) =
\delta_{n_1^0 0}$ leading to vanishing of the peak for $n_2^0=1$.

Next, we study the behavior of the system in the presence of an
applied AC field of amplitude $A$ and frequency $\omega_A$. In the
presence of such a field $\omega (\tau)= \omega_0 + A \sin(\omega_A
\tau)/\omega_A$. In Fig.\ \ref{fig6}(a), we show the behavior of the
peaks of $I^{\rm DC}_s$ as a function of $\omega_0$ for a fixed
$\omega_A=0.2$ and $A = 0.1$ in the presence of a constant magnetic
field. The peaks in $I^{\rm DC}_s$ occur at $\omega_0=0.4,0.6,0.8,1$
(from left to right); each of these correspond to two sets of
$(n_1^0,n_2^0)= (3,1)\, {\rm and}\, (-2,0),\, (2,1)\,{\rm and}\,
(-3,0),\, (1,1)\, {\rm and}\, (-4,0)$ and $(0,1)\,{\rm and}\,
(-5,0)$ respectively as predicted in Eq.\ \ref{condcons2}. In Fig.\
\ref{fig6}(b), we investigate the behavior for $I^{\rm DC}_s$ for a
periodically varying magnetic field as a function of $\omega_1$ for
$\omega_0=1.2$ and for same amplitude and frequency of the AC field.
We find several peaks in $I^{\rm DC}_s$; each of these peaks
corresponds to a fixed set of integers $(n_1^0, n_3^0)$ (Eq.\
\ref{condper2} with $\omega_2=0$) as shown in Table\ \ref{table1}.

\begin{table}
\begin{center}
\large
\begin{tabular}{?{0.25mm} c |  c |c ?{0.25mm} c| c| c ?{0.25mm}}
\hlinewd{1pt} $\omega_1$ & $n_1^0$  & $n_3^0$ &$\omega_1$ & $n_1^0$
& $n_3^0$\\ \hlinewd{1pt}

1.2 & 0 & 1 & 0.3 & 0 &  4  \\
\hline
1 & -1  & 1 &0.28 & 1 &  5\\
\hline
0.8 & 2  & 2 &0.24 & 0 &  5 \\
\hline
0.6 & 0  & 2 &0.2 & -3 &  3 \\
\hline
0.4 & -2  & 2 &0.2 & 0 &  6 \\
\hline
0.4 & 0  & 3 &0.08 & 0 & 15 \\
\hlinewd{1pt}
\end{tabular}
\end{center}
\caption {Tabulated values of $n_1^0$ and $n_3^0$ for all the peaks
that appear in Fig.\ \ref{fig7}(b) at specific $\omega_1$ values
listed above. Note that $n_2^0$ does not appear in the table since
the peaks correspond to $\omega_B=0$ so that their position are
independent of $n_2^0$ (Eq.\ \ref{condper2}).} \label{table1}
\end{table}

\begin{figure}[ht]
\centering
\includegraphics[width=4.2cm,height=3.2cm]{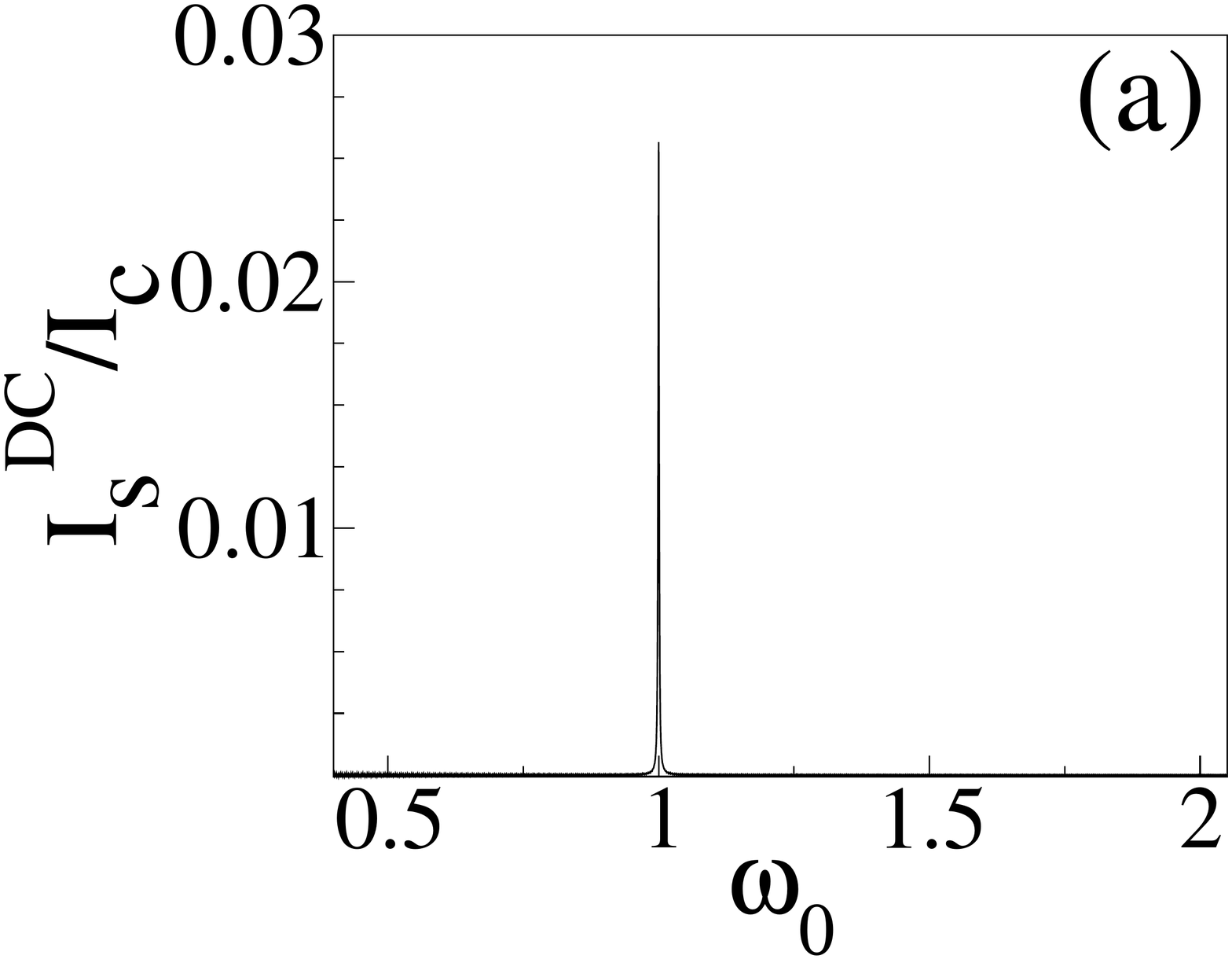}
\includegraphics[width=4.2cm,height=3.cm]{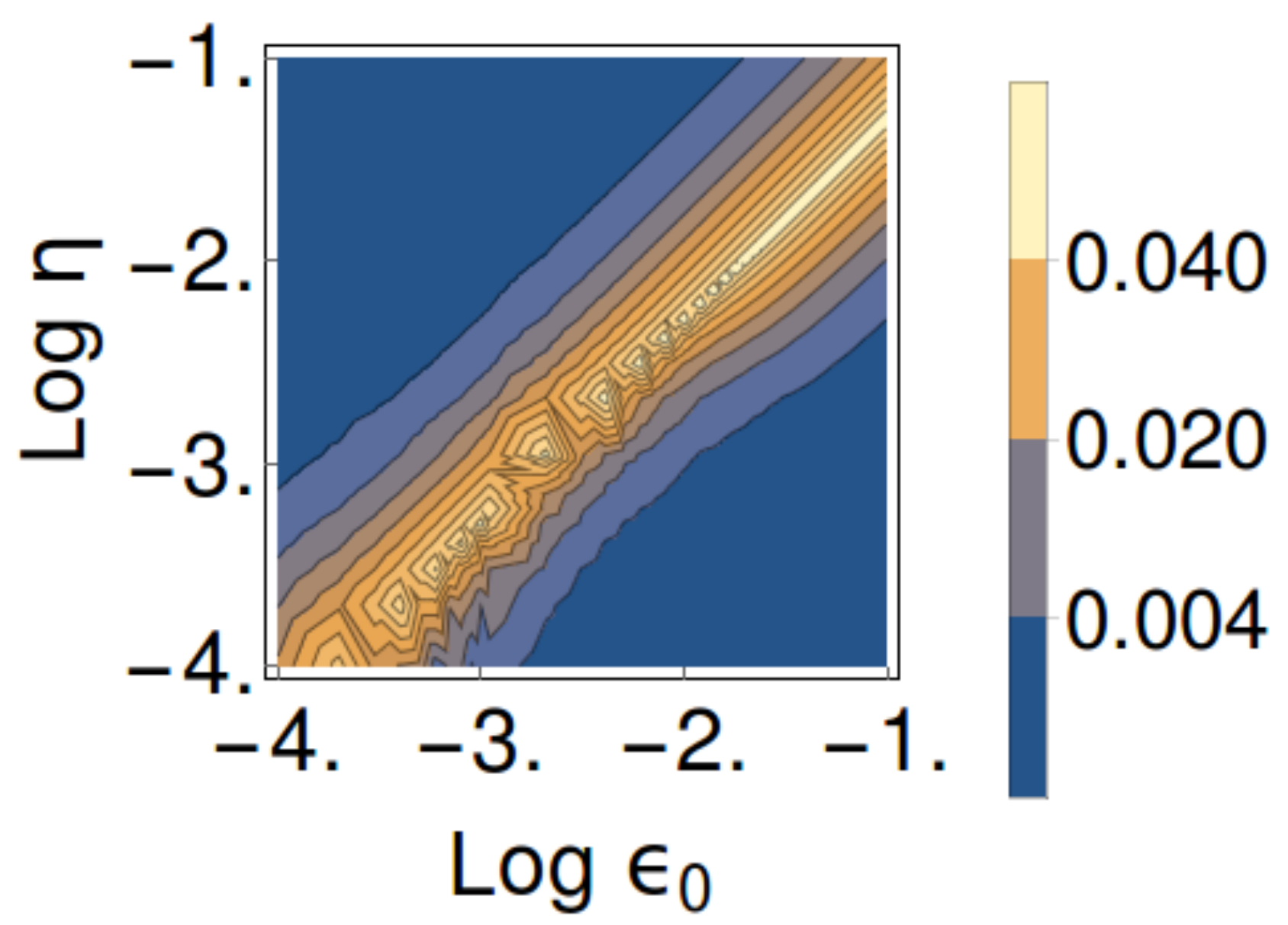}
\caption{(a)Plot of $I_s^{\rm DC}/I_c$ for a constant applied
magnetic field as a function of $\omega_0$ with $\eta=0.0001$. All
other parameters are same in Fig.\ \ref{fig4}(a). (b) Variation of
the peak height (for $n^0=1$) as a function of $\log \epsilon_0$ and
$\log \eta$ showing the presence of a line in the $\epsilon_0-\eta$
plane for which the peak height is maximal.} \label{fig7}
\end{figure}
\begin{figure}[ht]
\centering
\includegraphics[width=4.2cm,height=3.2cm]{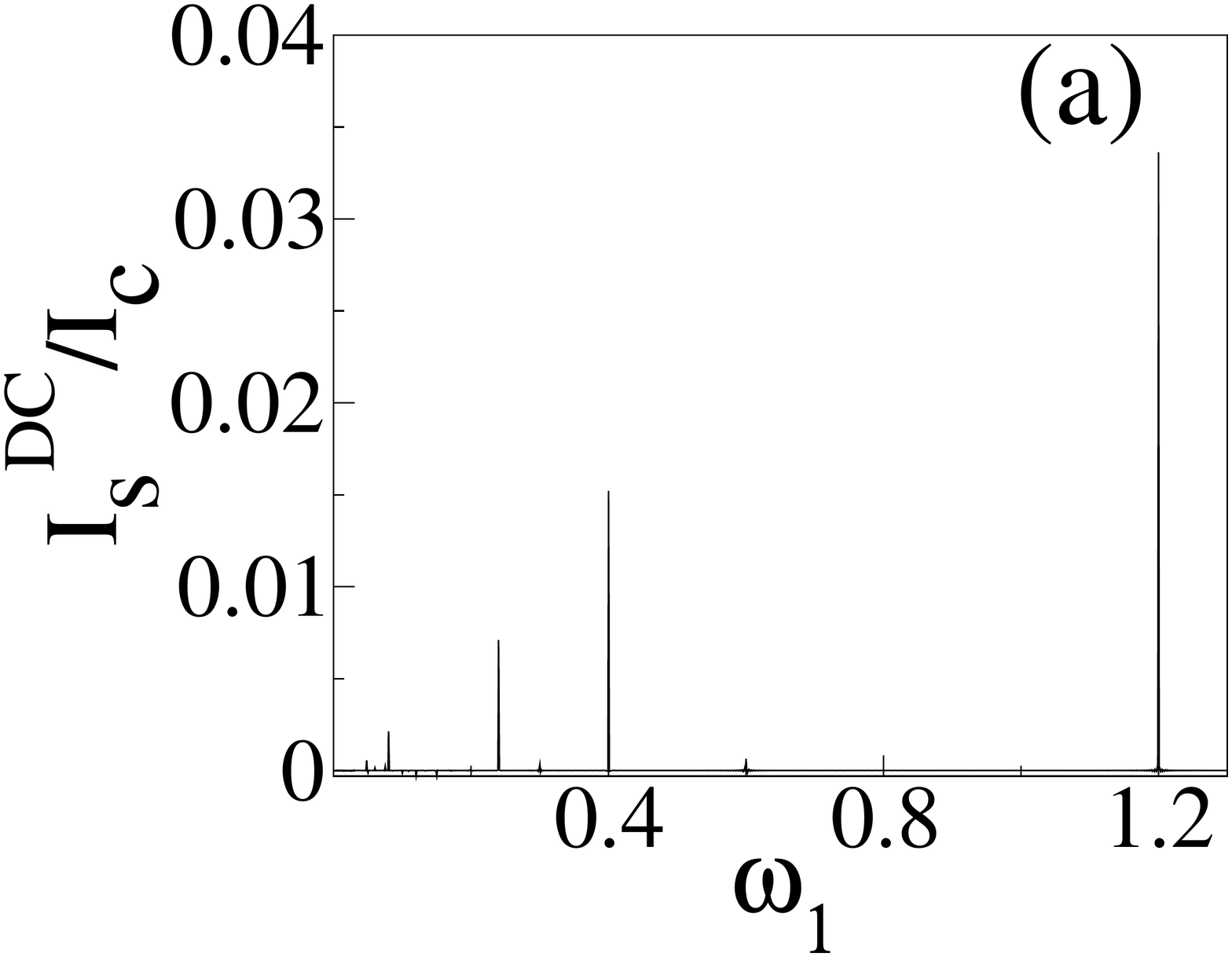}
\includegraphics[width=4.2cm,height=3.cm]{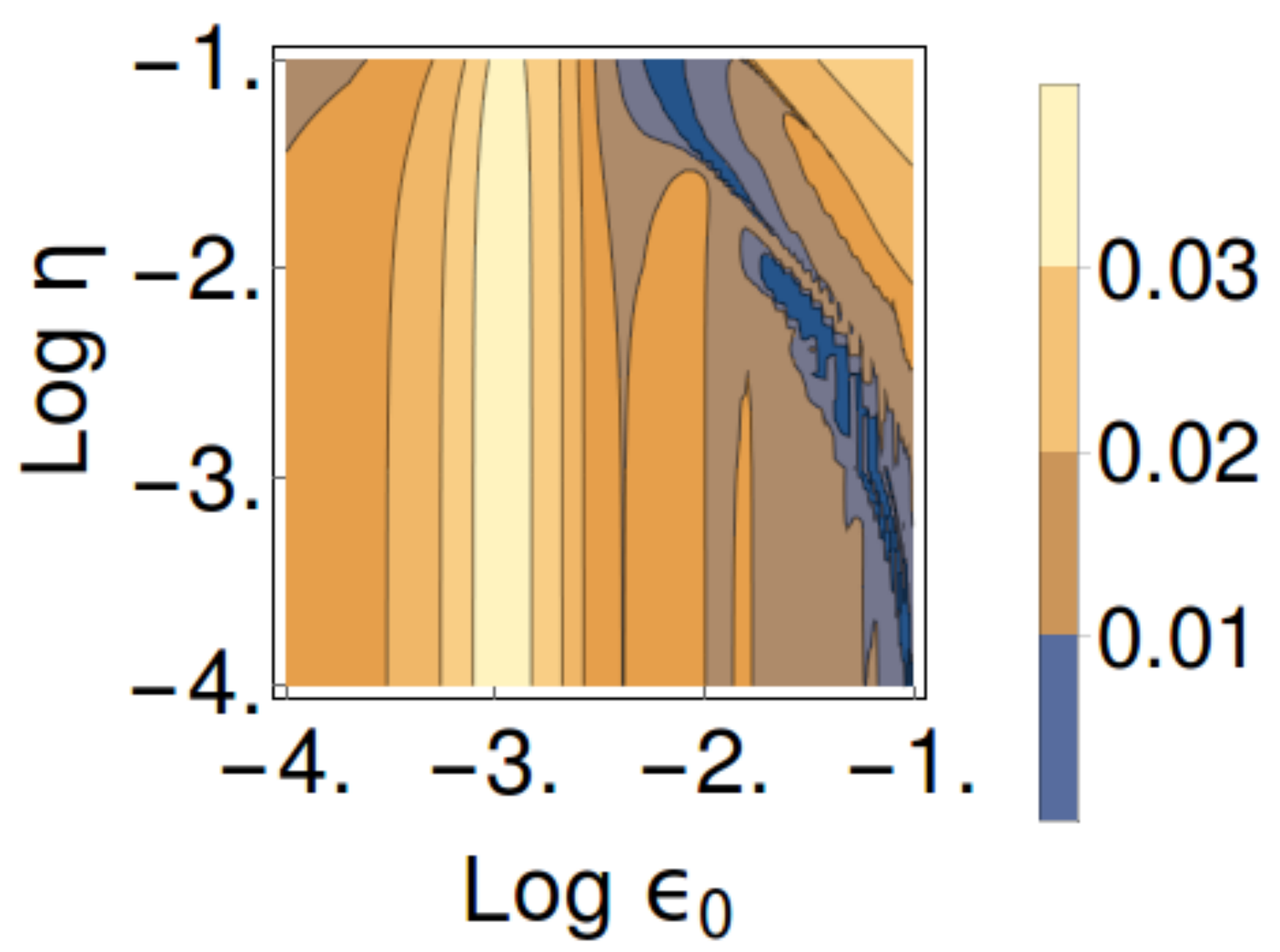}
\caption{(a)Plot of $I_s^{\rm DC}/I_c$ for a periodically varying
magnetic field as a function of $\omega_1$ with $\eta=0.0001$. All
other parameters are same in Fig.\ \ref{fig5}(b).(b) Variation of
the peak height (for $n_0=1$) as a function of $\log \epsilon_0$ and
$\log \eta$ showing the region of maximal peak height.} \label{fig8}
\end{figure}

\begin{figure}[ht]
\centering
\includegraphics[width=4.2cm,height=3.2cm]{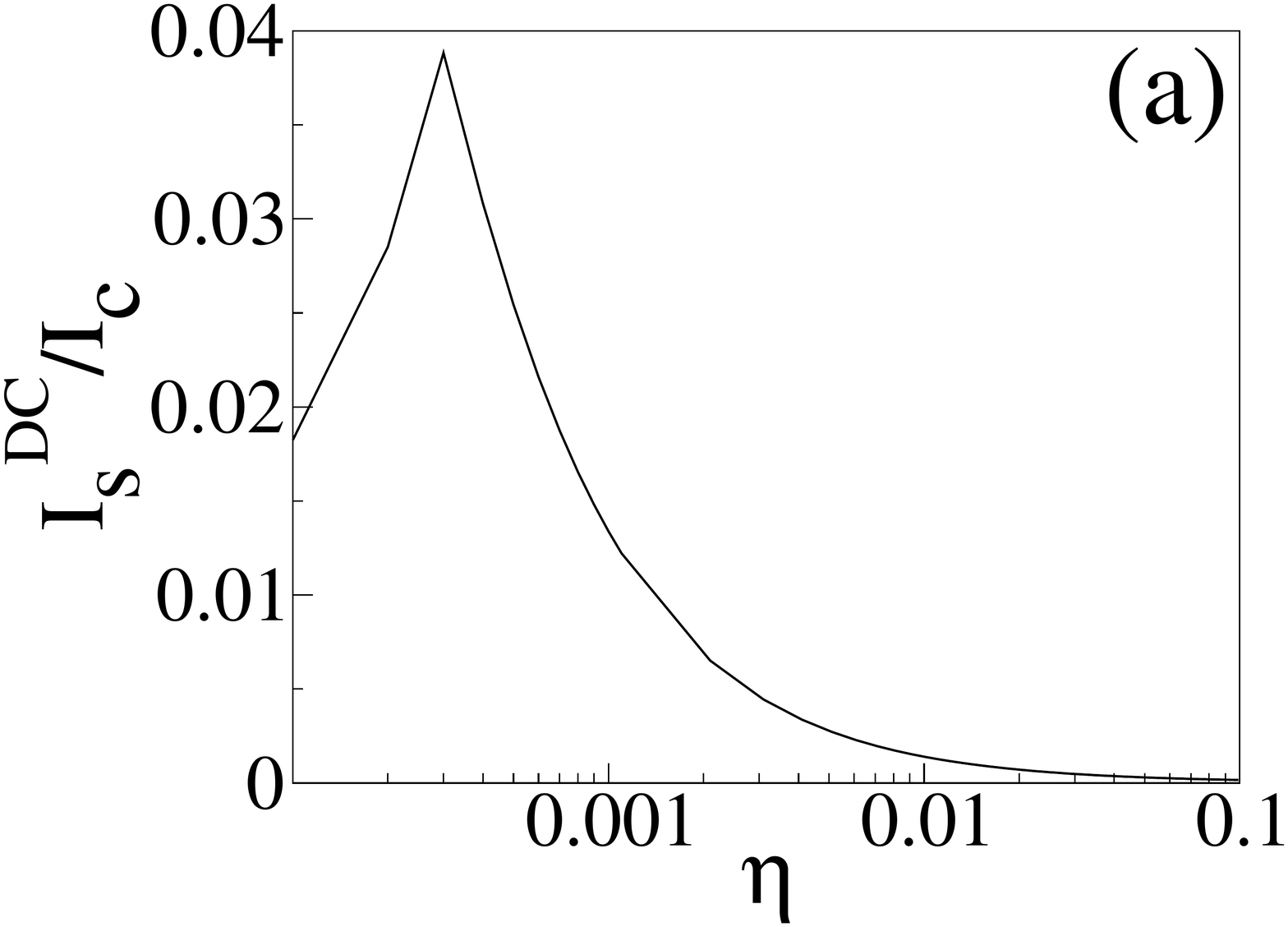}
\includegraphics[width=4.2cm,height=3.cm]{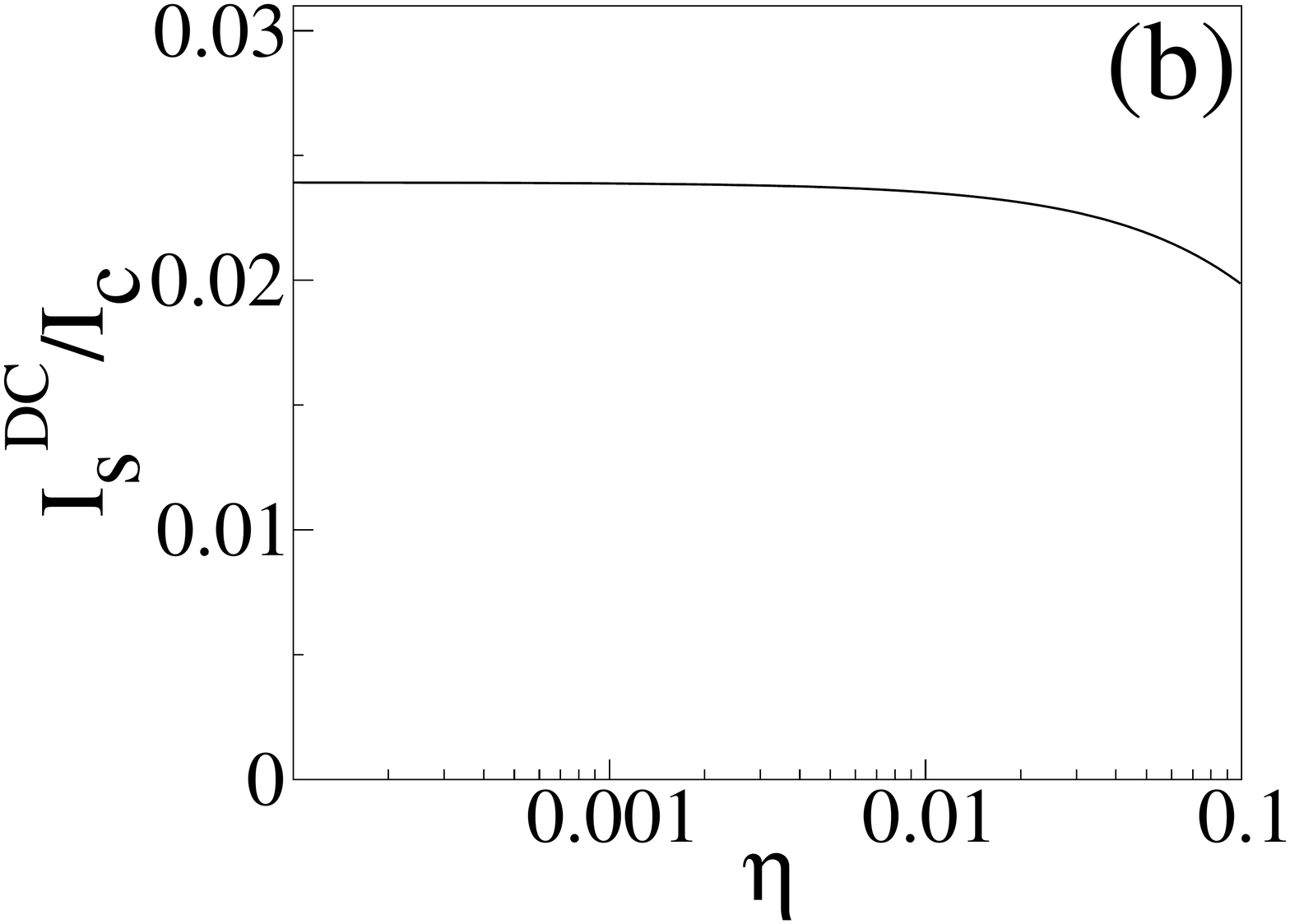}
\caption{(a)Plot of $I_s^{\rm DC}/I_c$ for a constant magnetic field
$\omega_0=1.0$ as a function of $eta$ with $\epsilon_0=0.0004$.(b) Similar plot
for periodic magnetic field with $\omega_1=1.2$. All
other parameters are same in Fig.\ \ref{fig4}.} \label{fig9}
\end{figure}

Next, we study the effect of dissipation on these peaks by plotting
$I^{\rm DC}_s$ as a function of $\omega_0$ in Fig.\ \ref{fig7}(a)
(for constant magnetic field) and as a function of $\omega_1$ in
Fig.\ \ref{fig8}(a) (periodic magnetic field) for $\eta=0.0001$. As
seen in both cases, the position of the peaks remain same as that
for $\eta=0$ in accordance with the analysis of Sec.\ \ref{gilres}.
The variation of the peak height as a function of $\log \epsilon_0$
and $\log \eta$ is shown in Figs. \ref{fig7}(b) for a constant
magnetic field. We find that the maximal peak-height occur along a
line in the $\epsilon_0-\eta$ plane. This can be seen more
clearly by plotting $I^{\rm DC}_s/I_c$ as a function of $\eta$ for a
fixed $\epsilon_0$ as shown in Fig.\ \ref{fig9}(a); the figure
displays a clear peak in $I_s^{\rm DC}$ at $\epsilon_0 \simeq
\eta$. This can be understood from Eq.\ \ref{thetadsol} and
\ref{pertt1} as follows. For the constant magnetic field, $z(\tau)=
\omega_c \tau$; consequently for small $\eta$, the correction to the
zeroth order solution from the dissipative term varies linearly with
$\eta$ (Eq.\ \ref{thetadsol})
\begin{eqnarray}
\delta \theta_d (\tau) \simeq \delta \theta(\tau) +  \theta_0 +
\sin(2\theta_0) \eta \omega \tau + ... \label{constmagdis}
\end{eqnarray}
where the ellipsis indicate higher order term in $\eta$. This
correction has opposite sign to the $\tau$-linear correction terms
(terms corresponding to $n=n^0 \mp 1$ in Eq.\ \ref{pertt1}) arising
due to a finite $\epsilon_0$ in $\delta \theta(\tau)$. The
corrections from $\eta$ and $\epsilon_0$ with opposite signs cancel
along some specific line $\epsilon_0-\eta$ plane leading to enhanced
better stability of the zeroth order solution and hence enhanced
peak height. We note that the angle of this line depends on details
of the relative magnitude of the correction terms. Thus we find that
the presence of dissipation in a nanomagnet may lead to enhancement
of the Shapiro-like steps for constant magnetic fields.

In contrast, as shown in \ref{fig8}(b), the peak height is almost
independent of $\eta$ for small $\eta$ for periodically varying
magnetic field. This can also be clearly seen from Fig.\
\ref{fig9}(b) where $I_s^{\rm DC}$ is shown to be indepedent of
$\eta$ for small $\eta$ at fixed $\epsilon_0$. For such fields,
$z(\tau)= \omega_2 \tau + \sin(\omega_1 \tau)/\omega_1$, where
$\omega_2 = \gamma_g K M_2/B_1 \ll \omega_1$ for our choice of
parameters. In this case, one can write, for small $\eta$
\begin{eqnarray}
\delta \theta_d (\tau) \simeq \delta \theta(\tau) +  \theta_0 +
\sin(2\theta_0) \eta \sin(\omega_1 \tau)/\omega_1+ ...
\label{constmagdis}
\end{eqnarray}
where the ellipsis indicate higher order term in $\eta$. Thus the
correction term is bounded and provides an oscillatory contribution
to $\theta(\tau)$. For small $\eta$, it is insignificant compared to
the correction term from $\epsilon_0$ and hence the peak height
stays almost independent of $\eta$. Thus we find that the role of
dissipation is minimal for small $\eta$ in case of periodically
varying magnetic fields. The oscillatory variation of the peak
height as a function of $\epsilon_0$ for a fixed $\eta$ can be
traced to its dependence on product of three Bessel functions as can
be seen from Eq.\ \ref{percorr}.

\begin{figure}[ht]
\centering
\includegraphics[width=4.2cm,height=3.2cm]{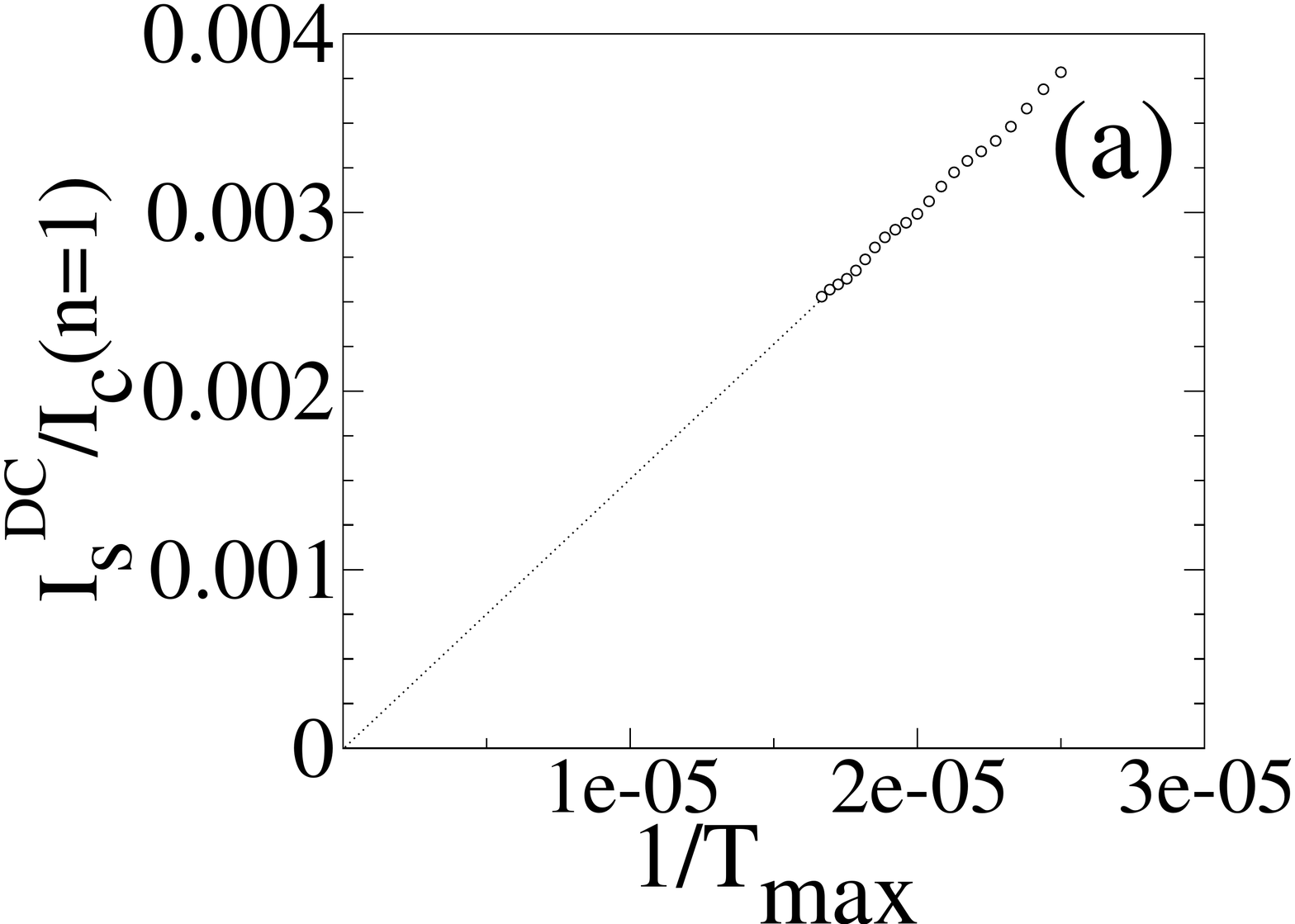}
\includegraphics[width=4.2cm,height=3.2cm]{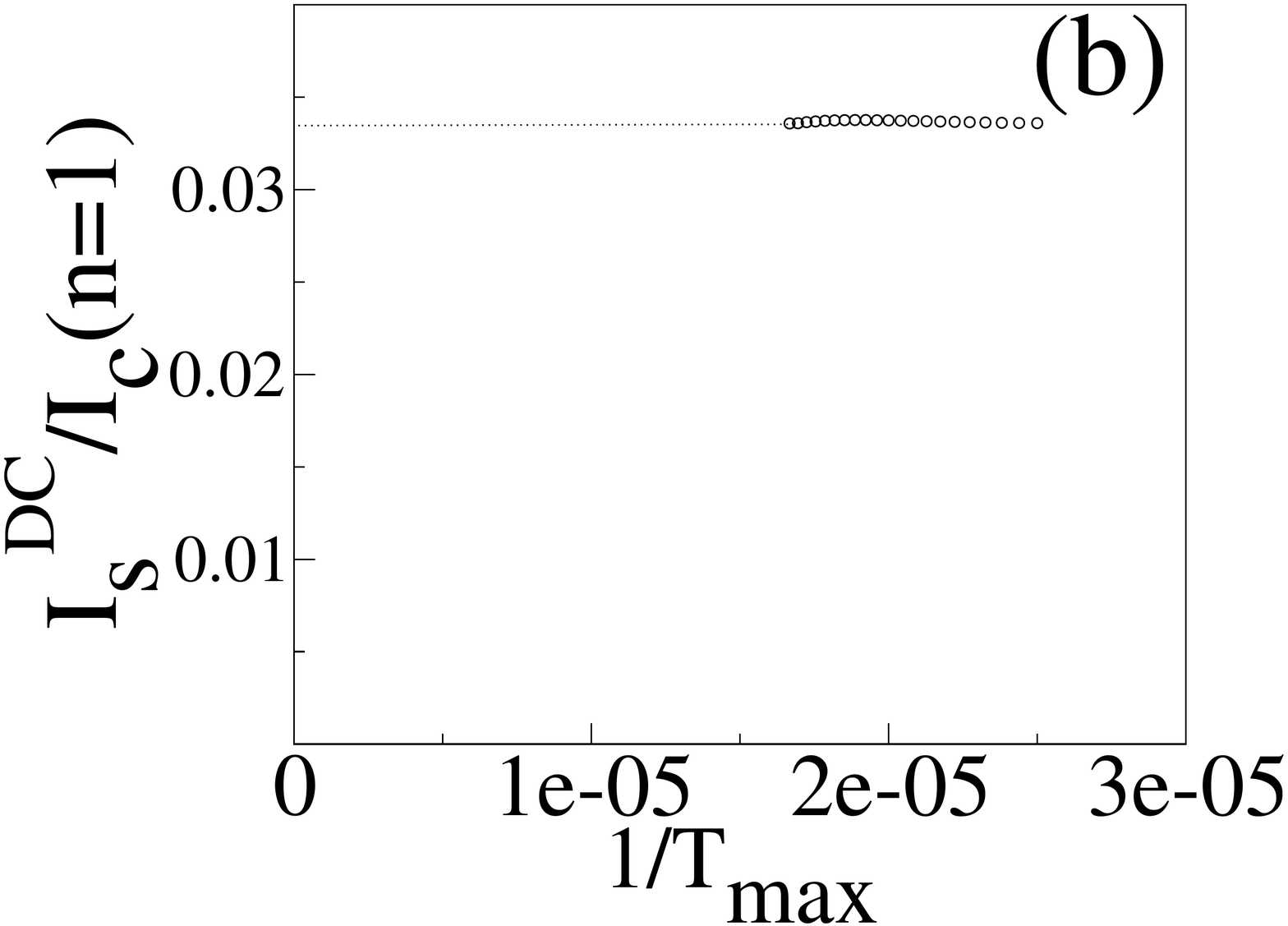}
\includegraphics[width=4.2cm,height=3.2cm]{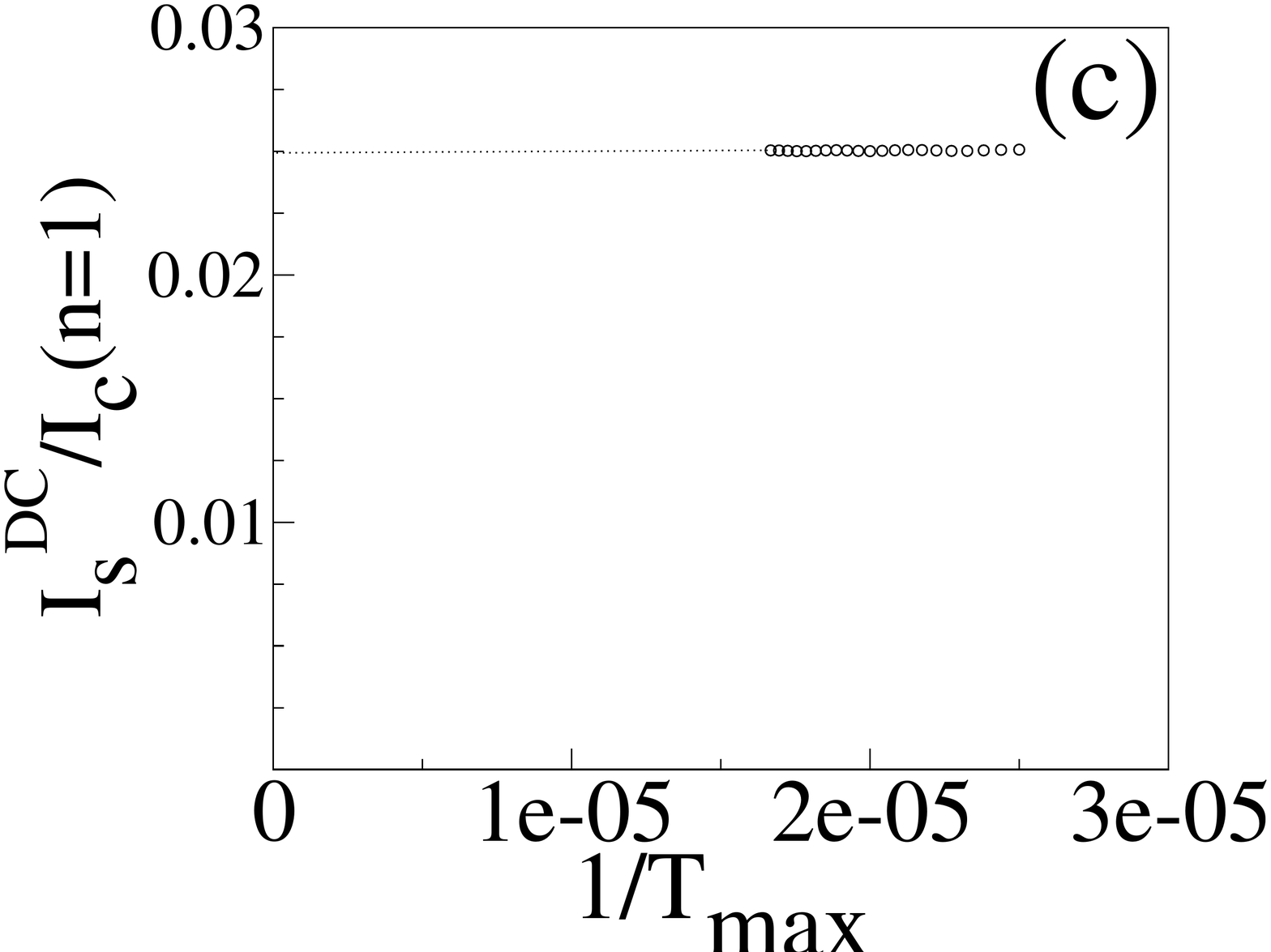}
\includegraphics[width=4.2cm,height=3.2cm]{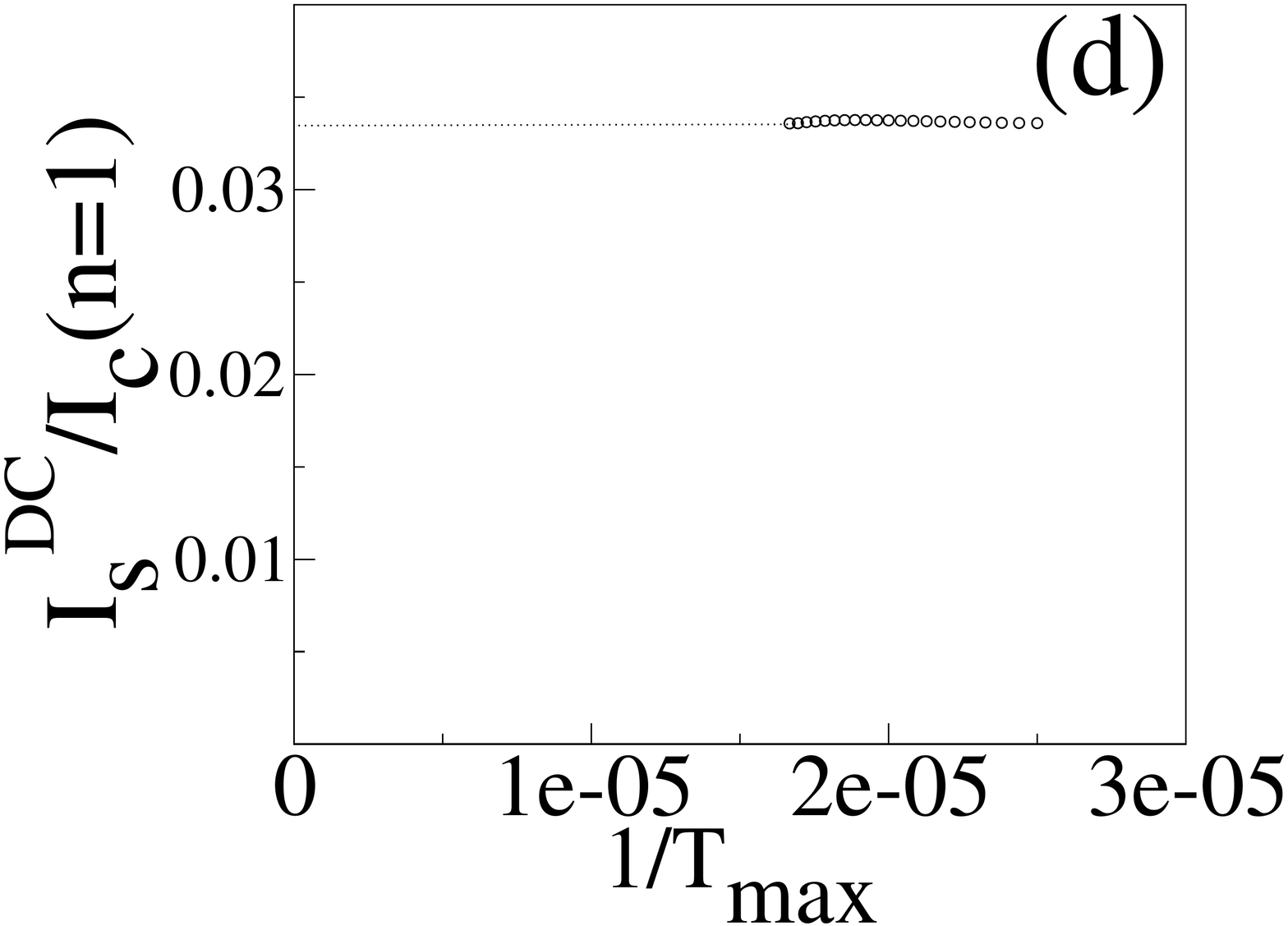}
 \caption{Plot $I_s^{\rm DC}/I_c$ as a function of $1/T_{\rm max}$
for (a) constant magnetic field with $\epsilon_0=0.001$ and
$\eta=0$, (b) periodically varying magnetic field with
$\epsilon_0=0.001$ and $\eta=0$, (c) constant magnetic field with
$\epsilon_0=\eta=0.001$ and (d) Periodically varying magnetic field
with $\epsilon_0=\eta=0.001$. All other parameters are same as in
Figs.\ \ref{fig4} and \ref{fig5}.} \label{fig10}
\end{figure}

Finally, we briefly study the effect of increasing $T_{\rm max}$ in
our numerical study. The relevance of this lies in the fact that for
any finite $\epsilon_0$ and $\eta$, our analytical results hold till
$\tau \sim T'$ (constant magnetic field) or $\tau \sim T'_p$
(periodic magnetic field) while the DC signal receives contribution
from all $T$. Thus it is necessary to ensure that these deviations
do not lead to qualitatively different results for the DC response.
To this end, we plot the height of the peak value of $I_{s}^{DC}$ as
a function of $1/T_{\rm max}$ in Fig.\ \ref{fig10}. We find from
Fig.\ \ref{fig10}(a) that for constant magnetic field, the peak
height indeed extrapolates to zero indicating that the Shapiro steps
will be destabilized due to perturbative corrections if $I_s$ is
averaged over very long time. However, we note from Fig.\
\ref{fig10}(c), $I_s^{\rm DC}$ could retain a non-zero value in the
presence of a finite dissipation parameter $\eta$. This could be
understood since the effect of damping, as shown in Fig.\
\ref{fig8}, negates that of $\epsilon_0$ on the peak value of
$I_s^{\rm DC}$. Furthermore, from Figs.\ \ref{fig10}(b) and (d), we
note that for the periodic magnetic fields the extrapolated value of
$I_s^{\rm DC}$ is a finite which is lead to finite Shapiro steps in
the I-V characteristics of theses JJs. Thus we expect that the
Shapiro-step like features in the I-V characteristics of the JJ to
be much more stable for periodically varying magnetic fields.

\section{Discussion}
\label{dissec}

In this work, we have studied a coupled JJ-nanomagnet system and
analyzed the behavior of the supercurrent in the JJ in the presence
of a finite coupling to the nanomagnet. We have provided a
perturbative analytical solution to both the LL and the LLG
equations governing the magnetization dynamics of the nanomagnet for
arbitrary time-dependent magnetic field applied along the easy axis
of the nanomagnet in the presence of weak coupling to the JJ and for
weak dissipation. We have estimated the regime of validity of our
perturbative results. We note here that whereas we have mainly
focussed on the dynamics of the critical current of the junction in
the presence of the nanomagnet in this work, the dynamics of the
nanomagnet itself may also have interesting features
\cite{buzhdin1,yury1}; we leave this issue for future work.

Using these results, we have studied the behavior of the
supercurrent of the JJ for constant and periodically varying
magnetic fields. The reason for choice of such magnetic fields are
that they are the only ones which lead to a fixed DC component of
$I_s(t)$ which in turn leads to Shapiro step-like features in the
I-V characteristics of a voltage-biased JJ. We note that while such
features are known for constant magnetic field from earlier works
\cite{cai1,buzhdin1}, the presence of such peaks in $I_s^{\rm DC}$
has not been theoretically reported for periodically time-varying
magnetic fields. Moreover, we show, both from our analytical results
and by performing exact numerics which supports these results, that
the peaks in $I_{s}^{\rm DC}$ for periodically varying magnetic
field are much more robust against increase of both $\epsilon_0$ and
$\eta$ compared to their constant field counterparts; we therefore
expect such peaks to be more experimentally accessible. We have also
studied the behavior of such JJ-nanomagnet systems in the presence
of external AC voltage. The presence of such a voltage leads to more
peaks in $I_{s}^{\rm DC}$ whose positions are accurately predicted
by our theoretical analysis. We note that our analysis, which
is carried out at zero temperature, is expected to be valid at low
temperature where $k_B T \ll \Delta_0, g \mu_B B$ so that the the
presence of thermal noise can be neglected. However, we point out
that the effect of such noise term in our formalism can be addressed
by adding a (white) noise term in the Gilbert equations following
standard procedure \cite{noise1}.

Our analysis could be easily extended to unconventional
superconductors hosting Majorana end states. For these junctions,
the current-voltage relation is $4 \pi$ periodic and given by $I_s =
I_c \sin (\gamma(t)/2)$ \cite{kit1,kwon1, kiril1}. An analysis using
this I-V relation immediately reveals that the Shapiro steps will be
present for $\omega_0= 2n^0 \omega_c$ for constant magnetic field
(Eq.\ \ref{condcons1}) and $\omega_0 = 2(n_2^0 \omega_1+n_1^0
\omega_2)$ for periodic magnetic fields (Eq.\ \ref{condper1}). The
additional factor of $2$ is a consequence of $4\pi$ periodicity
mentioned above. Thus coupling such JJs with Majorana end modes to
nanomagnets in the presence of a magnetic field may lead to new
experimental signatures of such end modes.

The experimental verification our work would involve preparing a
voltage biased JJ-nanomagnet system with sufficiently small values
of $\epsilon_0$. The current in such a junction, assuming a
resistive junction,  is given by
\begin{eqnarray}
I(t) = I_c \sin(\phi(t)-k_0 m_z) + V_0/R  + \frac{\hbar k_0}{2 e R}
dm_z/dt \label{vbias}
\end{eqnarray}
where $V_0$ is the bias voltage, $R$ is the resistance of the
junction and $\phi(t) = 2 e V_0 t/\hbar = \omega_0 t$. Thus the DC
component of the current will show additional spikes when Eqs.\
\ref{condcons1} (constant magnetic field) or \ref{condper1}
(periodic magnetic field) is satisfied. We note that it is essential
to have a voltage bias to observe these steps. This can be seen from
the fact that for current-biased junctions, the phase $\phi(t)$ is
not locked to a fixed value of $\omega_0 t$ but has to be obtained
from the solution of
\begin{eqnarray}
I = \frac{\hbar}{2eR} \partial_t \phi(t)  + I_c \sin[\phi(t) - k_0
m_z] - \frac{\hbar k_0}{2 e R} dm_z/dt, \label{cbias}
\end{eqnarray}
where $I$ is the bias current. Thus $\phi$ becomes a function of
$m_z$; we have checked both using perturbative analytic methods
\cite{kiril1} and exact numerics that in this case no steps exist.
This situation is to be contrasted to the case of standard Shapiro
steps induced via external radiation of amplitude $A$ where steps
can be shown to exist for both current and voltage biases
\cite{likharev1, kiril1}. Thus the present system would require a
voltage-biased junction for observation of Shapiro steps.

For experimental realization of such a system, this we envisage a 2D
thin film superconducting junction in the $x-y$ plane coupled to the
nanomagnet as shown in Fig.\ \ref{fig1}. We note that value of
Josephson energy and superconducting gap in a typical niobium film
are $E_J \sim 40$K and $\Delta_0 \sim 3$ meV respectively. Thus for
a typical magnetic field $\simeq 100$ Gauss, for which $\gamma_g
B_1= 0.28$GHz, one could estimate an $\epsilon_0 \simeq 0.0005$ for
$k_0 \simeq 0.1$. The Larmor frequency $\omega_L$ associated with
such magnetic field would be order of GHz while the spin-flip
processes responsible for any change in the Josephson current in
niobium junctions would be valid only if $\omega_L  \ge
\Delta_0/\hbar \sim 4.5$THz conforming the regime for our
perturbative analysis to hold. Moreover the critical current in such
junctions is $I_c = 2 e E_J / \hbar \simeq 1.5 \mu$A. Thus the peaks
in DC would correspond to $I_s^{\rm DC} \simeq 10$nA which is well
within detection capability of standard experiments. Typically such
peaks would require a voltage bias of $0.1 \mu$V which is also well
within present experimental capability. We also note here that such
experiments should also be possible with 1D junctions which have
been prepared experimentally in recent times using nanowires with
spin-orbit coupling \cite{rokhin1}.

To conclude, we have provided a perturbative analytic results for
supercurrent of a coupled JJ-nanomagnet system in the limit of weak
coupling between them and in the presence of a time dependent field
applied to the system. Using this analytic result and exact
numerical solution of the LL and the LLG equations, we predict
existence of peaks in $I_s^{\rm DC}$ for both constant and periodic
magnetic fields which are expected to provide Shapiro-like steps in
the I-V characteristics of a voltage-biased JJ without the presence
of external AC drive. We have analyzed the effect of finite
dissipation of the nanomagnet and the presence of external AC drive
on these peaks and discussed experiments which can test our theory.

{\it Acknowledgement:} KS acknowledges DST/RFBR grant
INT/RUS/RFBR/P-249. RG acknowledges SPM fellowship from CSIR for
support. The reported study was partially funded by RFBR according
to the research projects 15-29-01217 and 16-52-45011, India.


\begin{thebibliography}{99}

\bibitem{likharev1} K. K. Likharev, Rev. Mod. Phys. {\bf 51}, 101
(1979); K. K. Likharev, {\it Dynamics of Josephson Junctions and
Circuits}, (Taylor and Francis, London, 1986).

\bibitem{kit1} A. Kitaev, Phys. Usp. {\bf 44}, 131 (2001).

\bibitem{kwon1} H-J Kwon, K. Sengupta, V.M. Yakovenko, Eur. Phys. J. B
{\bf 37}, 349 (2004).

\bibitem{jay1} R. M. Lutchyn, J. D. Sau, and S. Das Sarma, Phys. Rev. Lett.
{\bf 105}, 077001 (2010); Y. Oreg, G. Refael, and F. von Oppen,
Phys. Rev. Lett. {\bf 105}, 177002 (2010).

\bibitem{gil1} L. Fu and C. L. Kane, Phys. Rev. Lett. {\bf 100}, 096407
(2008). [9] L. Fu and C. L. Kane, Phys. Rev. B {\bf 79}, 161408(R)
(2009); J. D. Sau, R. M. Lutchyn, S. Tewari, and S. Das Sarma, Phys.
Rev. Lett. {\bf 104}, 040502 (2010); J. Alicea, Phys. Rev. B {\bf
81}, 125318 (2010); A. Cook and M. Franz, Phys. Rev. B {\bf 84},
201105(R) (2011); J. D. Sau and S. D. Sarma, Nat. Commun. {\bf 3},
964 (2012); A. Das, Y. Ronen, Y. Most, Y. Oreg, M. Heiblum, and H.
Shtrikman, Nat. Phys. {\bf 8}, 887 (2012); M. T. Deng, C. L. Yu, G.
Y. Huang, M. Larsson, P. Caroff, and H. Q. Xu, Nano Lett. {\bf 12},
6414 (2012); A. D. K. Finck, D. J. Van Harlingen, P. K. Mohseni, K.
Jung, and X. Li, Phys. Rev. Lett. {\bf 110}, 126406 (2013); H. O. H.
Churchill, V. Fatemi, K. Grove-Rasmussen, M. T. Deng, P. Caroff, H.
Q. Xu, and C. M. Marcus, Phys. Rev. B {\bf 87}, 241401 (2013); M.
Cheng and R. M. Lutchyn, Phys. Rev. B {\bf 92}, 134516 (2015).

\bibitem{rokhin1} L. P. Rokhinson, X. Liu, and J. K. Furdyna, Nat. Phys. {\bf 8}, 795
(2012); V. Mourik, K. Zuo, S. M. Frolov, S. R. Plissard, E. P. A. M.
Bakkers, and L. P. Kouwenhoven, Science {\bf 336}, 1003 (2012); W.
Chang, V. E. Manucharyan, T. S. Jespersen, J. Nyg�ard, and C. M.
Marcus, Phys. Rev. Lett. {\bf 110}, 217005 (2013); S. Nadj- Perge,
I. K. Drozdov, J. Li, H. Chen, S. Jeon, J. Seo, A. H. MacDonald, B.
A. Bernevig, and A. Yazdani, Science {\bf 346}, 602 (2014); E. J. H.
Lee, X. Jiang, M. Houzet, R. Aguado, C. M. Lieber, and S. D.
Franceschi, Nat. Nanotechnol. {\bf 9}, 79 (2014).

\bibitem{shapiro1} S. Shapiro, Physical Review Letters {\bf 11}, 80
(1963).

\bibitem{evenlit}M. Houzet, J. S.Meyer, D.M. Badiane, and
L. I. Glazman, Phys. Rev. Lett. {\bf 111}, 046401 (2013); L. Jiang,
D. Pekker, J. Alicea, G. Refael, and Y. Oreg, and F. von Oppen,
Phys. Rev. Lett. {\bf 107}, 236401 (2011); F. Domnguez, F. Hassler,
and G. Platero, Phys. Rev. B {\bf 86}, 140503 (2012); D. I. Pikulin
and Y. V. Nazarov, Phys. Rev. B {\bf 86}, 140504(R) (2012); J. D.
Sau, E. Berg, and B. I. Halperin, arXiv:1206.4596 (unpublished).

\bibitem{kiril1} M. Maiti, K. M. Kulikov, K. Sengupta, and
Yu. M. Shukrinov, Phys. Rev. B {\bf 92}, 224501 (2015).

\bibitem{commentsub} We note that, as shown in Ref.\
\onlinecite{kiril1}, odd subharmonic steps can occur in these
junction for a wide range of dissipation parameter in the presence
of finite junction capacitance.

\bibitem{time1} A. Ardavan, O. Rival, J. J. L. Morton, S. J. Blundell, A. M. Tyryshkin,
G. A. Timco, and R. E. P. Winpenny, \prl {\bf 98}, 057201 (2007).

\bibitem{loss1} M. N. Leuenberger, and D. Loss, Nature {\bf 410}, 789 (2001).

\bibitem{ref1} A. R. Rocha, V. M. Garc\'{\i}a-Su\'{a}rez, S. W. Bailey, C. J. Lambert,
J. Ferrer AND S. Sanvito, Nat. Mat. {\bf 4}, 335 (2005).

\bibitem{ref2} L. Bogani, and W. Wernsdorfer, Nature Mater. \textbf{7}, 179 (2008).

\bibitem{tunnel} J.R. Friedman, M. R. Sarachik, J. Tejada, and R. Ziolo, Phys. Rev. Lett. {\bf 76}, 3830 (1996);
L. Thomas,  F. Lionti, R. Ballou, D. Gatteschi, R. Sessoli, and  B.
Barbara, Nature {\bf 383}, 145 (1996).

\bibitem{qi} M. Trif, F. Troiani, D. Stepanenko, and D. Loss, Phys. Rev. Lett. 101, 217201 (2008);
P. Santini, S. Carretta, F. Troiani, and G. Amoretti, {\it ibid}
107, 230502 (2011).

\bibitem{ent1} W. Wernsdorfer, N. Aliaga-Alcalde, D.N Hendrickson, and
G. Christou, Nature {\bf 416}, 406 (2002); S. Hill, R. S. Edwards,
N. Aliaga-Alcalde, and G. Christou, Science {\bf 302}, 1015 (2003);
G. Timco et. al., Nat. Nanotech. {\bf 4}, 173 (2009).

\bibitem{spin1} R. Caciuffo, G. Amoretti, A. Murani, R. Sessoli, A. Caneschi, and
D. Gatteschi, Phys. Rev. Lett. {\bf 81}, 4744 (1998); L.  Mirabeau,
M.  Hennion,  H.  Casalta, H.  Andres,  H.U.  G\"{u}del, A.V.
Irodiva, and A. Caneschi, {\it ibid} {\bf 83}, 628 (1999).

\bibitem{spin2} M. L. Baker et. al. , Nat. Phys. {\bf 8}, 906911 (2012).

\bibitem{spin3} E. Burzur\'{\i}, A. S. Zyazin, A. Cornia, and H. S. J. van der Zant, \prl {\bf 109}, 147203 (2012).

\bibitem{spin4} B. Abdollahipour, J. Abouie, and N. Ebrahimi, AIP Advances {\bf 5}, 097156 (2015).

\bibitem{kulik1} I. O. Kulik, Zh. Eksp. Teor. Fiz. {\bf 49}, 585 (1966)
[Sov. Phys. JETP {\bf 22}, 841 (1966)].

\bibitem{bula1}L. N. Bulaevskii, V. V. Kuzii, and A. A. Sobyanin, Zh. Eksp.
Teor. Fiz. {\bf 25}, 314 (1977) [JETP Lett. {\bf 25}, 290 (1977)].

\bibitem{nuss1} J.-X. Zhu, Z. Nussinov, A. Shnirman, and A. V. Balatsky, Phys.
Rev. Lett. {\bf 92}, 107001 (2004); Z. Nussinov, A. Shnirman, D. P.
Arovas, A. V. Balatsky, and J. X. Zhu, Phys. Rev. B {\bf 71}, 214520
(2005).

\bibitem{nazarov1} C. Padurariu and Yu. V. Nazarov, Phys. Rev. B {\bf 81}, 144519
(2010).

\bibitem{delan1} L. Dell�Anna, A. Zazunov, R. Egger, and T. Martin, Phys.
Rev. B {\bf 75}, 085305 (2007).

\bibitem{cai1} L. Cai and E.M. Chudnovsky, Phys. Rev. B {\bf 82}, 104429 (2010);
L. Cai, D. A. Garanin, and E. M. Chudnovsky, Phys. Rev. B, {\bf 87},
024418 (2013).

\bibitem{buzhdin1} A. I. Buzdin, Rev. Mod. Phys. {\bf 77}, 935
(2005); A. Buzdin, Phys. Rev. Lett. {\bf 101}, 107005 (2008); F.
Konschelle and A. Buzdin, Phys. Rev. Lett. {\bf 102}, 017001 (2009).

\bibitem{yury1} Y. M. Shukrinov, I. R. Rahmonov, K. Sengupta, and A.
Buzdin, Appl. Phys. Lett. {\bf 110}, 182407 (2017).

\bibitem{xav1} X. Waintal and P. Brower, Phys. Rev. B {\bf 65},
054407 (2002).

\bibitem{linder1} I. Kulagina and J. Linder, Phys. Rev. B {\bf 90}, 054504
(2014); J. Linder and T. Yokoyama Phys. Rev. B {\bf 83}, 012501
(2011).

\bibitem{holm1} C. Holmqvist, S. Teber, and M. Fogelstrom, Phys. Rev. B {\bf 83},
104521 (2011).

\bibitem{wernsdorfer1}W. Wernsdorfer, Adv. Chem. Phys. {\bf 118}, 99
(2001); W. Wernsdorfer, Supercond. Sci. Technol. {\bf 22}, 064013
(2009).

\bibitem{thirion1} C. Thirion, W. Wernsdorfer, and D. Mailly, Nat. Mater. {\bf 2}, 524
(2003).

\bibitem{gil1} T. L. Gilbert, IEEE Transactions on Magnetics {\bf 40}, 3443
(2004).

\bibitem{comment1} We note that both the LL and the LLG equations
used here can be derived using Euler-Lagrange method from standard
Lagrangians for arbitrary time-dependent magnetic fields. See Ref.\
\onlinecite{gil1} for details.


\bibitem{balat1} J-X Zhu and A. V. Balatsky, \prb {\bf 67}, 174505
(2003).

\bibitem{commentmag} For such a time dependent magnetic field along
$\hat y$ the accompanying electric field as obtained from $\nabla
\times \vec E = - \partial B/\partial t$ may be chosen to be along
$\hat z$; thus it would not couple to a JJ located in the x-y plane.

\bibitem{noise1} P-W. Ma and S. L. Dudarev \prb {\bf 86}, 054416 (2012);
U. Atxitia, O. Chubykalo-Fesenko, R. W. Chantrell, U. Nowak, and
A. Rebei, Phys. Rev. Lett. {\bf 102}, 057203 (2009); T. Bose and S.
Trimper, Phys. Rev. B {\bf 81}, 104413 (2010); W. Coffey, Yu. P.
Kalmykov, and J. T. Waldron, {\it The Langevin Equation}, 2nd ed.
(World Scientific, Singapore, 2004).




\end{thebibliography}
\end{document}